\documentclass{aa} 

\usepackage{graphicx} 
\usepackage{txfonts} 
\usepackage{natbib} 
 
\newcommand{\Mpc}{$h^{-1}$\thinspace Mpc}

\newcommand{\be}{\begin{equation}}
\newcommand{\ee}{\end{equation}}

\def\apj{ApJ} 
\def\apjl{ApJL} 
\def\apjs{ApJS} 
\def\aj{AJ} 
\def\aap{A\&A} 
\def\mnras{MNRAS}

\usepackage{color}
\usepackage{amstext}

\begin{document}    
 
\title{Correlation function:   biasing and fractal properties  of the cosmic web}
 
\author{J. Einasto\inst{1,2,3} 
\and  G. H\"utsi\inst{4} 
\and  T. Kuutma\inst{1}
\and   M. Einasto\inst{1}  
}
 
\institute{Tartu Observatory, University of Tartu, 61602 T\~oravere, Estonia 
\and 
Estonian Academy of Sciences, 10130 Tallinn, Estonia
\and  
ICRANet, Piazza della Repubblica, 10, 65122 Pescara, Italy 
\and
National Institute of Chemical Physics and Biophysics, 
Tallinn 10143, Estonia 
} 
 
\date{ Received 07/02/2020; accepted 08/06/2020}  
 
\authorrunning{Einasto et al.} 
 
\titlerunning{Correlation function and the cosmic  web} 
 
\offprints{J. Einasto, e-mail: jaan.einasto@ut.ee} 
 
\abstract {}
{Our goal is to determine how the spatial correlation function of galaxies
  describes biasing and fractal properties  of the cosmic web. }
{We calculated spatial correlation functions of galaxies, $\xi(r)$,
  structure functions, $g(r)=1 +\xi(r)$, gradient functions,
  $\gamma(r)= d \log g(r)/ d \log r$, and fractal dimension functions,
  $D(r)= 3+\gamma(r)$, using dark matter particles of the biased $\Lambda$
  cold dark matter (CDM) simulation, observed galaxies of the Sloan
  Digital Sky Survey (SDSS), and simulated galaxies of the Millennium
  and EAGLE simulations.  We analysed how these functions describe
  fractal and biasing properties of the cosmic web.}
{ The correlation functions of the biased $\Lambda$CDM model samples
  at small distances (particle and galaxy separations), $r \le 2.25$~\Mpc,
  describe the distribution of matter inside dark matter (DM) halos. In real and simulated
  galaxy samples, only the brightest galaxies in clusters are visible, and
  the transition from clusters to filaments occurs at a distance
  $r \approx 0.8 - 1.5$~\Mpc.  At larger separations, the correlation
  functions describe the distribution of matter and galaxies in the
  whole cosmic web.  The effective fractal dimension of the cosmic web
  is a continuous function of the distance (separation).  Real and simulated
  galaxies of low luminosity, $M_r \ge -19$, have almost identical
  correlation lengths and amplitudes, indicating that dwarf galaxies
  are satellites of brighter galaxies, and do not form a smooth
  population in voids.}
{
  The combination of several  physical processes { (e.g.} the formation
of halos along the caustics of particle trajectories and the
phase synchronisation of density perturbations on various
scales) transforms the initial random density field to
the current highly non-random density field.  Galaxy formation
is suppressed in voids, which increases the amplitudes of correlation functions and
power spectra of galaxies, and increases the large-scale bias
parameter. The combined evidence leads to the large-scale bias
parameter of $L_\star$ galaxies the value $b_\star  =1.85 \pm 0.15$.
We find     $r_0(L_\star) = 7.20
\pm 0.19$ for  the correlation length of $L_\star$ galaxies. }

\maketitle

\section{Introduction} 

{ Early data on the distribution of galaxies came from the counts  of
  galaxies on the sky.  The available data suggested that the distribution
  of galaxies on the sky is dominated by an almost random field population
  with randomly spaced clusters.  The distribution can be described by
  the two-point correlation function of galaxies
  \citep{Peebles:1980aa}.  The two-point correlation is a function of
  distance and describes the excess probability of finding two
  galaxies separated by this distance. It is not sensitive to the
  pattern of the cosmic web because it measures only the amplitude of
  the density fluctuations of the density field of galaxies, not the phase
  information, which is essential to describe the cosmic web.  To
  investigate the structure of the cosmic web in detail, various other
  methods have been developed. For a recent overview, see
  \citet{VandeWeygaert:2016zt}.  In spite of these limitations, large
  areas in the study of galaxy distribution exist where 
  the two-point correlation function and its Fourier companion, the power
  spectrum of density perturbations, are useful.

  One area in which the correlation functions can be applied is the
  biasing phenomenon.  \citet{Peebles:1975} showed that the two-point
  correlation function of galaxies can be represented as a power law
  $\xi(r) = (r/r_0)^{\gamma}$, where $r_0 \approx 5$~\Mpc\ is the
  correlation length, and $\gamma \approx -1.7$ is the slope of the
  correlation function.  The bias parameter $b$ can be defined as the
  ratio of the amplitudes of the galaxy correlation functions (or power spectra $P$)
to the mass, $b^2 = P_g/P_m$.  The analyses of
  galaxy correlation functions by \citet{Lahav:2002aa},
  \citet{Verde:2002aa}, and \citet{Zehavi:2005aa} suggested that the
  large-scale bias parameter is $b \approx 1$.  A similar result was
  found from the analysis of power spectra of galaxies by
  \citet{Tegmark:2004aa} and \citet{Zehavi:2011aa}.

In contrast, \citet{Einasto:1993ab, Einasto:1999aa} showed that the
bias parameter is inversely proportional to the fraction of matter in
galaxies, $b = 1/F_g$, which means that $b > 1$.
\citet{Mandelbaum:2013aa} used luminous red giant (LRG) galaxies of
the Sloan Digital Sky Survey (SDSS) to  obtain a value of $b \approx 2$ for the large-scale
bias.  \citet{Klypin:2016aa} used several
MultiDark simulations and determined the bias factor of power spectra of
halos to the power spectra of mass, $b= 1.95$.  \citet{Einasto:2019aa}
compared power spectra of biased samples of a $\Lambda$ cold dark
matter (CDM) model with the full dark matter (DM) sample of the same
model. The authors found the value $b_{\ast}= 1.85 \pm 0.15$ for the
bias parameter of the model equivalent of 
$L_{\ast}$ galaxies.  

The reason for this large difference in bias parameter estimates was
not clear.  In order to understand what causes the differences between
various determinations of bias parameters, we need to
analyse the technique of the correlation function analysis in more detail.

The main goal of this paper is to understand the biasing
phenomenon as the relation between distributions of matter and
galaxies by applying the two-point correlation function.  In doing so, we
determine why analyses of correlation functions and
power spectra of galaxies yield very different values for the bias
parameter $b$.

The slope of the correlation function $\gamma$ contains information on
the fractal character of the cosmic web.  Early data suggested that
a power law with fixed $\gamma$ is valid over a very wide range of
distances.  It is currently well accepted that galaxies form in DM-dominated halos
\citep{White:1978}.  Thus the spatial correlation function at small
separations $r$ characterises the galaxy distribution within
DM-dominated halos or clusters, and on larger separations, the galaxy
distribution in the whole cosmic web. This leads to departures of the
correlation 
function from a pure power law. 
The study of fractal properties of the cosmic web using correlation
functions of various samples is the second goal of this paper.  The
relation between spatial and projected correlation functions is
investigated in a separate paper by \citet{Einasto:2020ab}.
}

Following \citet{Einasto:2019aa}, we use a
$\Lambda$CDM model as basis that is calculated in a box of size 512~\Mpc.  In
the simulations we select particles from the same set for biased samples
and for unbiased samples, and observational selection effects can be
ignored.  We take advantage of the fact that the positions
of all particles are known for this model.  This allows us to compare correlation
functions of dark matter and simulated galaxies differentially.

For the comparison, we analyse absolute magnitude (volume) limited
SDSS samples, as found by \citet{Liivamagi:2012aa} and
\citet{Tempel:2014ab}.  We also analyse the galaxy samples from the
Millennium simulation by \citet{Springel:2005aa} and the galaxy
catalogues of the EAGLE simulation provided by
\citet{McAlpine:2016aa}.

To investigate biasing and fractal  properties of the
cosmic web in more detail, we use the galaxy correlation function,
$\xi(r)$, and its derivatives: the structure function, 
$g(r)=1 +\xi(r)$, its log-log gradient, the gradient function,
$\gamma(r)= d \log g(r)/ d \log r$, and the fractal dimension function,
$D(r)= 3+\gamma(r) = 3+ d \log g(r)/ d \log r$.

The paper is organised as follows.  In the next section we describe
the simulation and the observational data we used, the methods for calculating the
correlation functions and their derivatives, and the first results of the
analysis of these functions for simulated and real galaxy samples of
various luminosity limits. { In section 3 we discuss the biasing properties
of LCDM and galaxy samples, and the luminosity dependence of
the correlation functions and their derivatives.}  In section 4 we study
the fractal properties of the cosmic web as described by the correlation
function and its derivatives.  The last section presents our
conclusions.

\section{Methods, data, and first results} 
 
In this section we describe the simulation and the observational data we used and 
the methods for calculating the correlation functions.  We start by
describing the methods for calculating the correlation and related functions.
Thereafter we describe the $\Lambda$CDM numerical simulation we used to
determine density fields at various density threshold levels to simulate
galaxy samples of different luminosity.  Next we describe our basic
observation sample selected from the Sloan Digital Sky Survey (SDSS)
and from the EAGLE and Millennium simulations, which we used for comparison with
observational samples. Finally, we describe our first results from
calculating the correlation functions.

\subsection{Calculation of  the correlation functions}

The natural estimator to determine the  two-point spatial correlation
function is 
\begin{equation}
  \xi_N(r) = {DD(r) \over RR(r)} - 1,
  \label{nat}
\end{equation}
and the \citet{Landy:1993ve} estimator is
\begin{equation}
 \xi_{LS}(r)= {DD(r)-2DR(r)+RR(r) \over RR(r)}.
  \label{LS}
\end{equation}
In these formulae, $r$ is the galaxy pair separation (distance), and $DD(r)$,
$DR(r),$ and $RR(r)$ are normalised counts of galaxy-galaxy,
galaxy-random, and random-random pairs at a distance $r$ of the pair members.

We applied the  \citet{Landy:1993ve}  estimator for the samples
of SDSS galaxies, and the natural estimator for the cubical EAGLE and
Millennium model galaxy samples.  The model samples have two scale
parameters, the length of the cube, $L$ in \Mpc, and the maximum
distance for calculating the correlation function, $L_{max}$.  The second
parameter is used to exclude pairs with separations larger than
$L_{max}$ in the loop over galaxies to count pairs.  This speeds up
the calculations considerably. The separation interval $r$ from
0 to $L_{max}$ was divided into 100 equal bins of length $L_{max}/100$
in case of the EAGLE and Millennium samples, and into $N_b$ equal bins of
length $\log L_{max}/N_b$ for the SDSS samples.  By varying $L_{max}$ and
$N_b,$ we can use different bin sizes that are optimised for the particular
sample.  The core of the Fortran program we used for the Millennium and EAGLE
samples was written previously to investigate galaxy correlation
functions by \citet{Einasto:1986ab}, \citet{Klypin:1989aa},
and \citet{Einasto:1997aa, Einasto:1997ab}. At these times, computers
were very slow and optimisation of programs was crucial. The
random-random pair count $RR(r)$ was found by applying the random number
generator {\tt ran2} by \citet{Press:1992gf}, using one or two million
particles. To speed up the calculation of random-random pairs, we split
random samples into ten independent subsamples, as was also done by
\citet{Keihanen:2019aa}.

$\text{The }\Lambda$CDM model samples contain all particles with local density
labels $\rho \ge \rho_0$.  To derive correlation functions of these
samples, a conventional method cannot by used because the number of particles
is too large, up to $512^3$ in the full unbiased model.  To determine the
correlation functions of the $\Lambda$CDM samples, we used
the \citet{Szapudi:2005aa} method.  Conventional methods scale as
$O(N^2)$.  When the number of objects $N$ becomes large, the method is
too slow. Measurements of power spectra can be obtained with fast
Fourier transforms (FFT), which scale as $O(N \log N)$.  The
\citet{Szapudi:2005aa} method also uses the FFT to calculate correlation
functions and scales as $O(N \log N)$.  The method is an
implementation of the algorithm {\tt eSpICE}, the Euclidean version of
{\tt SpICE} by \citet{Szapudi:2001aa}.  As input, the method uses
density fields on grids $1024^3$, $2048^3$, and $3072^3$.  Correlation
functions are found with $L_{max}=100$~\Mpc, and with 200, 400, and 600
linear bins, respectively.

In the following analysis we use the spatial correlation function,
$\xi(r)$, and the pair correlation or structure function,
$g(r) = 1+\xi(r)$, to characterise the galaxy distribution in
space; for details see \citet{Martinez:2002fu}.  
Instead of the slope of the correlation
function as a parameter, $\gamma$, we calculate the log-log gradient
of the pair correlation function as a function of distance,
\begin{equation}
  \gamma(r)= {d \log g(r) \over d \log r},
  \label{gamma}
\end{equation}
which we call gradient function, or simply  $\gamma(r)$ function. It is related to 
the effective fractal dimension $D(r)$ of  samples at mean separation of galaxies
at $r$,  
\begin{equation}
 D(r) =3+ \gamma(r).
  \label{dimens}
\end{equation}
We call the $D(r)=3+\gamma(r)$ function the fractal dimension
function.  \citet{Martinez:2002fu} defined the correlation dimension
$D_2 = 3 + d \log \hat{g}(r) / d \log r$, where $\hat{g}(r)$ is the
average of the structure function, $\hat{g}(r) = 1/V \int_0^r{g(r`) d
  V}$. For our study we prefer to use the local value of the structure
function to define its gradient.  
The effective fractal dimension of a random distribution of
galaxies is $D = 3$, and $\gamma =0$, respectively; in sheets, $D = 2$
and $\gamma = -1$; in a filamentary distribution, $D = 1$ and
$\gamma = -2$; and within clusters, $D = 0$ and $\gamma = -3$.

To determine the gradient function  $\gamma(r),$ we
use a linear fit of the structure function  $g(r)$.  At small distances, the
function changes very rapidly, and we use for its gradient at $r$ just
the previous and following $\log g(r)$ value divided by the log step size in
$r$.  At larger distances, we use a linear fit on the distance
interval in $\pm m$ steps from the particular $r$ value of $\log g(r)$,
presented as a table.  The fit is computed using the {\tt fit} subroutine
by \citet{Press:1992gf}, which gives the gradient and its error.
The correlation functions and their derivatives were calculated with a
constant linear or logarithmic step size.

{ The correlation functions, $\xi(r)$, are shown in Fig.~\ref{fig:Fig1},
  structure  functions, $g(r)=1 + \xi(r)$, in Fig.~\ref{fig:Fig2}, and
  the fractal dimension functions, $D(r) = 3+\gamma(r)$, in
  Fig.~\ref{fig:Fig3}, for all simulated and real galaxy samples.}  As
usual, we use the separation $r$, where the correlation function has
a unit value, $\xi(r_0) = 1.0$, as the correlation length of the sample 
$r_0$.

{ We calculated the relative correlation functions for all samples of
  LCDM models, which define the bias functions, 
\begin{equation}
  b(r, \rho_0) = \sqrt{\xi(r,\rho_0)/\xi(r,0)}.
\label{bias}  
\end{equation}
The bias functions have a plateau $6 \le r \le 20$~\Mpc, see
Fig.~\ref{fig:Fig7}.  This feature is similar to the plateau around
$k \approx 0.03$~$h$~Mpc$^{-1}$ of relative power spectra
\citep{Einasto:2019aa}.  We used this plateau to measure the amplitude
of the correlation function.  In this
separation range, the correlation functions are almost linear in log-log representation, therefore the exact
location of the reference point influences our results only quantitatively, but
makes little qualitative difference. 
 Our choice was $r_6=6$~\Mpc}.  We calculated for all
samples the value of the correlation function at distance (galaxy
separation) $r_6=6$~\Mpc, $\xi_6$; this parameter characterises the
amplitude of the correlation function.  It is complimentary to the
correlation radius $r_0$. { At smaller distances, the correlation functions
  are influenced by strong non-linear effects, and at larger distances, the
  correlation functions for highly biased samples have wiggles, which
  hampers the comparison of samples with various particle density limits}.

\subsection{Biased model samples}

The simulations of the evolution of the cosmic web were performed in a box
of size $L_0=512$~\Mpc, with a resolution $N_{\mathrm{grid}} = 512$ and
with $N_{\mathrm{part}} = N_{\mathrm{grid}}^3$ particles. The spectrum for the initial
density fluctuation was generated using the COSMICS code by
\citet{Bertschinger:1995},  { assuming concordance $\Lambda$CDM
cosmological parameters \citep{Bahcall:1999aa}:}
$\Omega_{\mathrm{m}} = 0.28$, $\Omega_{\Lambda} = 0.72$,
$\sigma_8 = 0.84$, and the dimensionless Hubble constant $h = 0.73$.
To generate the initial data, we used the baryonic matter density
$\Omega_{\mathrm{b}}= 0.044$ (\citet{Tegmark:2004}).  Calculations
were performed with the GADGET-2 code by \citet{Springel:2005}.  The
same model was used by \citet{Einasto:2019aa} to investigate the
biasing phenomenon using power spectra.

{\scriptsize 
\begin{table}[ht] 
\caption{LCDM particle-density-limited models.} 
\label{Tab1}                         
\centering
\begin{tabular}{lrlll}
\hline  \hline
Sample   & $\rho_0$&$b(\rho_0)$ &$r_0$ & $\xi_6$ \\  
\hline  
(1)&(2)&(3)&(4)&(5)\\ 
\hline  
LCDM.00   &  0.0 & 1.000  & 4.86 & 0.729\\
LCDM.05 &  0.5 & 1.143 &  5.75 & 0.953\\
LCDM.1   &  1.0 & 1.285 &  6.75 & 1.203\\
LCDM.2   &  2.0 & 1.447  & 7.83 & 1.530\\
LCDM.3   &  3.0 & 1.548  & 8.47 & 1.754\\
LCDM.4   &  4.0  &  1.621  & 8.89 & 1.924\\
LCDM.5   &  5.0 &  1.677  & 9.21  & 2.061\\
LCDM.7   &  7.5 &  1.778  & 9.77  &2.317\\
LCDM.10 &  10.0 & 1.849  & 10.14 &2.507\\
LCDM.15 &  15.0 & 1.952  & 10.76 &2.794\\
LCDM.20 &  20.0 & 2.031  & 11.20  &3.021 \\
LCDM.30 &  30.0 & 2.156  & 11.96 &3.402\\
LCDM.40 &  40.0 & 2.257  & 12.56 &3.725\\
LCDM.50 &  50.0 & 2.342  & 12.56 &3.725\\
LCDM.80 &  80.0 & 2.528  & 14.4  &4.662\\
LCDM.100&100.0& 2.626  & 14.9  &5.038\\
\hline 
\end{tabular} 
\tablefoot{
The columns show the
(1) sample name; 
(2)  the particle-density limit $\rho_0$;
(3) the bias parameter, calculated from correlation functions of biased models
with particle-density limits  $\rho_0$;
(4) the correlation length $r_0$ in \Mpc; and
(5) the correlation function amplitude $\xi_6$ at $r_6=6.0$~\Mpc.}
\end{table} 
} 

For all simulation particles and all simulation epochs, we calculated
the local density values at particle locations, $\rho$, using the positions
of the  27 nearest particles, including the particle itself.  Densities
were expressed in units of the mean density of the whole simulation.
{  We formed biased model  samples,  selected  on the particle density
at the present epoch.  The full $\Lambda$CDM model
includes all particles.  Following \citet{Einasto:2019aa}, we formed 
biased model samples that contained particles above a
certain limit, $\rho \ge \rho_0$, in units of the mean density of
the simulation.  As shown by \citet{Einasto:2019aa}, this sharp density
limit allows determining density fields of simulated galaxies whose
geometrical properties are close to the density fields of real galaxies. }
The biased samples are denoted LCDM.$i$, where $i$ denotes the
particle-density limit $\rho_0$.  The full DM model includes all
particles and corresponds to the particle-density limit
$\rho_0 = 0$, and therefore it is denoted as LCDM.00.  The main data
on the biased model samples are given in Table~\ref{Tab1}.  
We calculated for all samples the correlation length, $r_0$, and the amplitude 
of the correlation function at $r_6=6$~\Mpc, $\xi_6=\xi(6)$, 
all  as functions of the particle density limit, $\rho_0$.

\begin{figure*}[ht]
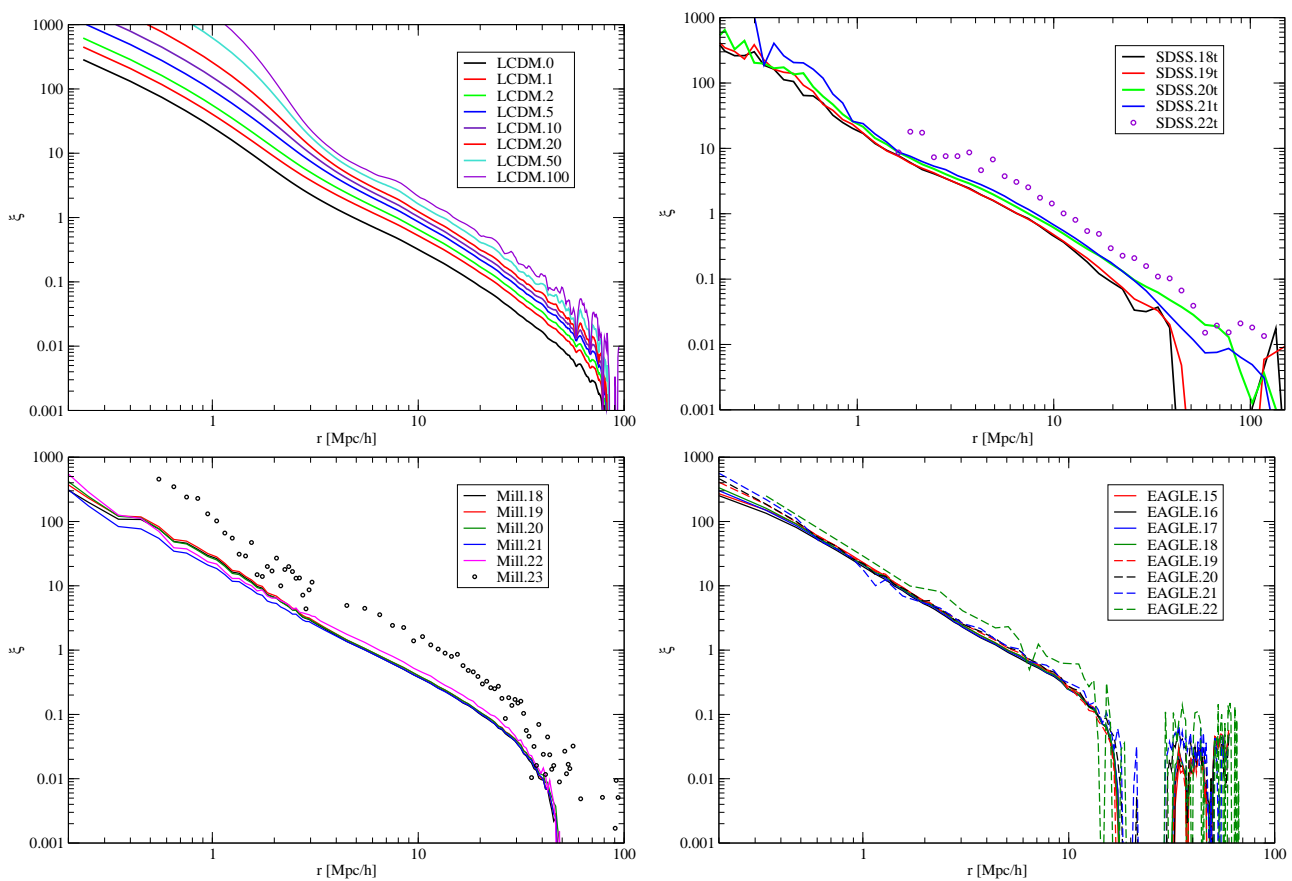
 
\centering 
\hspace{1mm}  
\resizebox{0.45\textwidth}{!}{\includegraphics*{LCDM-CF-rD.eps}}
\hspace{1mm}  
\resizebox{0.45\textwidth}{!}{\includegraphics*{cfSDSSthres.eps}}\\
\hspace{1mm}  
\resizebox{0.45\textwidth}{!}{\includegraphics*{cfmill_500_100-10Sum.eps}}
\hspace{1mm}  
\resizebox{0.45\textwidth}{!}{\includegraphics*{cf_100EAGLE_Thres.eps}}
\caption{Correlation functions of galaxies,
  $\xi(r)$. {\em The panels from top left to bottom right}
   show LCDM models for different particle selection limits; for
  SDSS galaxies using five luminosity thresholds; for six luminosity
  bins of Millennium galaxies, and for eight luminosity thresholds of
  EAGLE galaxies. }
\label{fig:Fig1} 
\end{figure*} 

\begin{figure*}[ht]
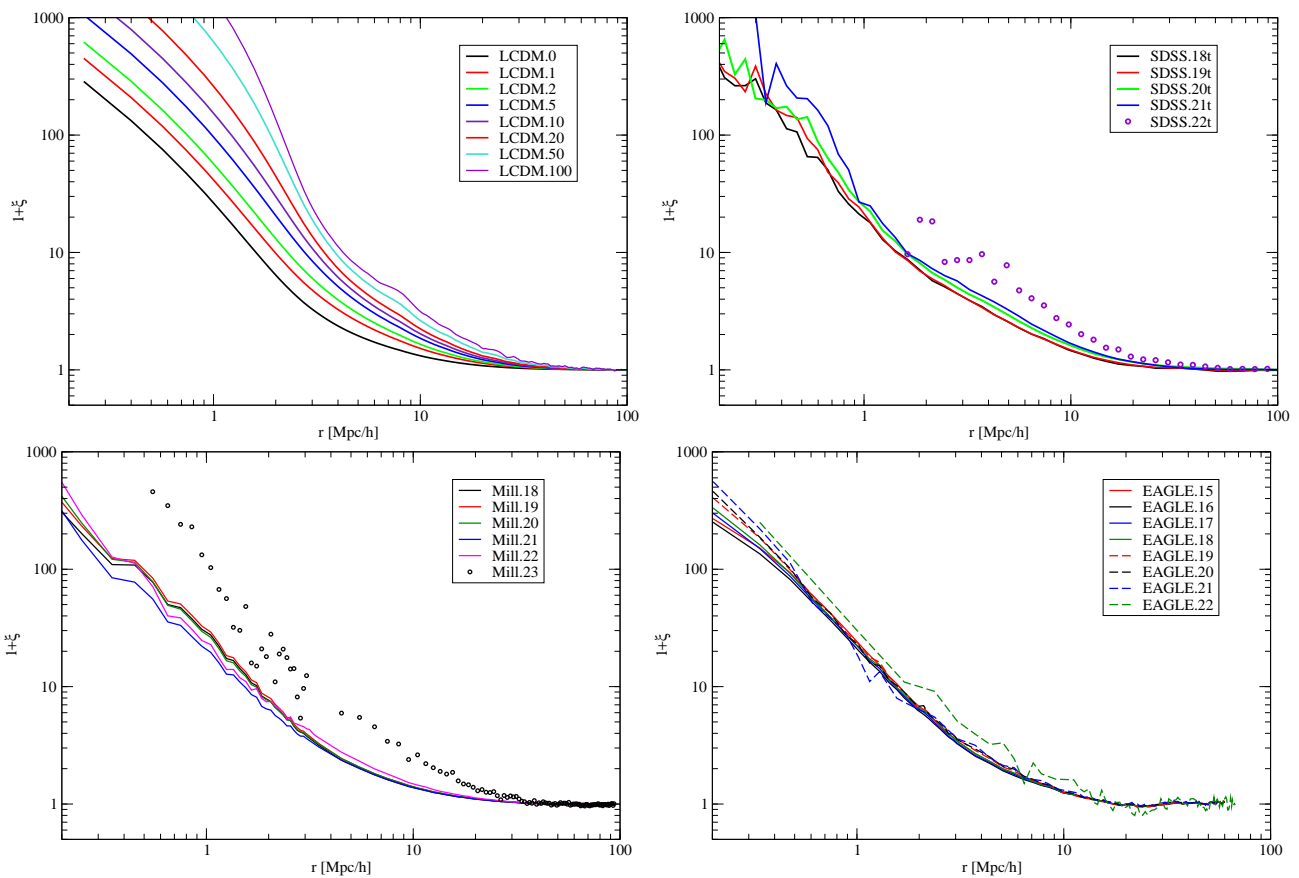
 
\centering 
\hspace{1mm}
\hspace{1mm}  
\resizebox{0.45\textwidth}{!}{\includegraphics*{LCDM-CFplus-rC.eps}}
\hspace{1mm}  
\resizebox{0.45\textwidth}{!}{\includegraphics*{cfPlusSDSSthres.eps}}\\
\hspace{1mm}  
\resizebox{0.45\textwidth}{!}{\includegraphics*{cfplusmill_500_100-10Sum.eps}}
\hspace{1mm}  
\resizebox{0.45\textwidth}{!}{\includegraphics*{cfPlus_100EAGLE_Thres.eps}}
\caption{Pair correlation or structure
  functions, $g(r)=1+\xi(r)$. The location of the panels is the same as in Fig.~{\ref{fig:Fig1}.}}
\label{fig:Fig2} 
\end{figure*}

\begin{figure*}[ht]
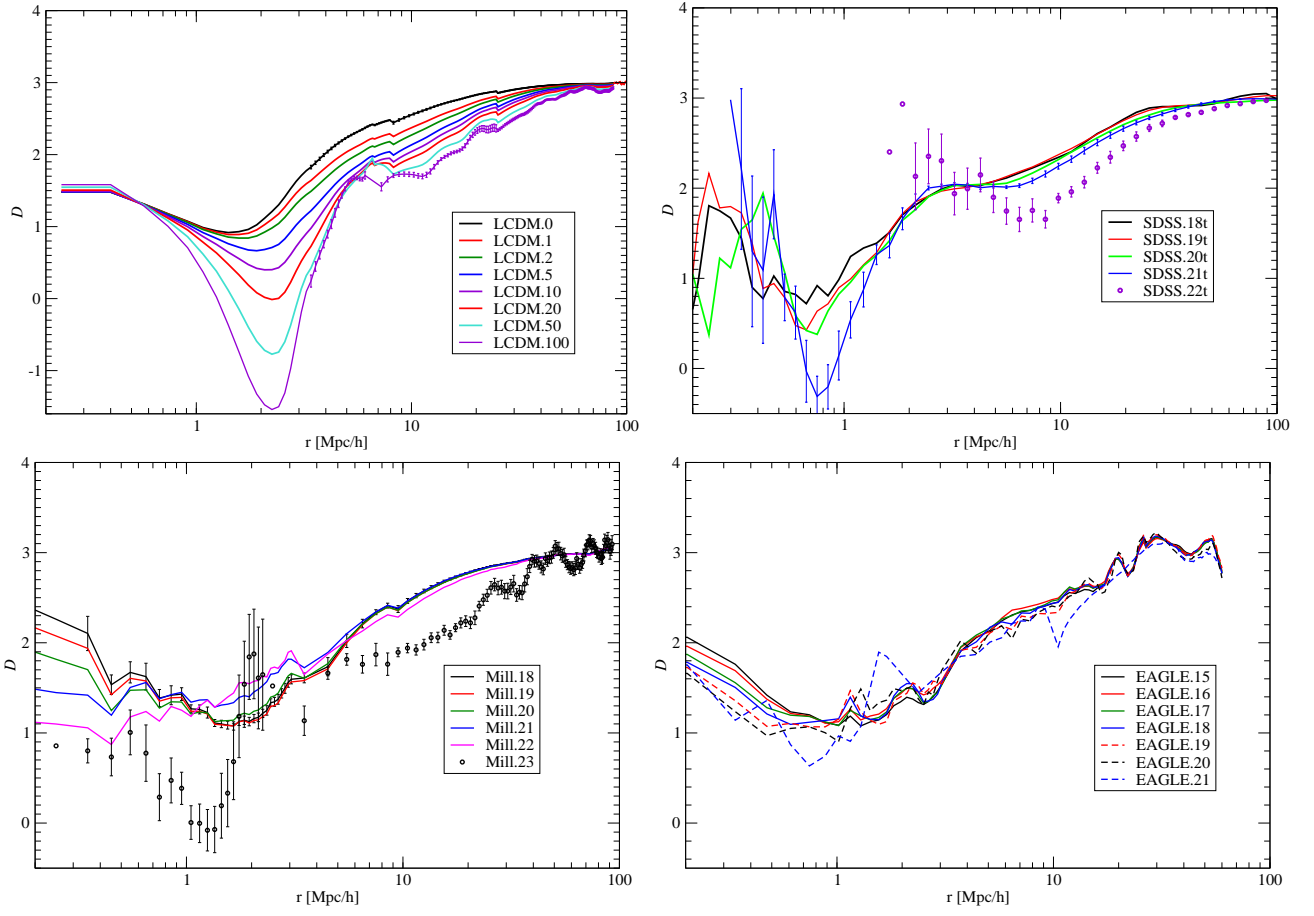
 
\centering 
\hspace{1mm}  
\hspace{1mm}  
\resizebox{0.45\textwidth}{!}{\includegraphics*{LCDM-gamma-rC.eps}}
\hspace{1mm}  
\resizebox{0.45\textwidth}{!}{\includegraphics*{SDSS-gamma-rMult.eps}}\\
\hspace{1mm}  
\resizebox{0.45\textwidth}{!}{\includegraphics*{cfmillGamma_500_10bin.eps}}
\hspace{1mm}  
\resizebox{0.45\textwidth}{!}{\includegraphics*{cfGamma_100EAGLE_Thres.eps}}
\caption{Fractal dimension functions, $D(r)=3+\gamma(r) $. The location of
the  panels is the same as in Fig.~{\ref{fig:Fig1}} Errors are found with the {\tt
    fit} procedure, and are given for several representative samples.}
\label{fig:Fig3} 
\end{figure*}

The gradient of the structure function changes very rapidly at small $r.$
  For this reason, we calculated correlation functions
using two resolutions, 3072 and 1024.  Fractal dimension functions
were calculated for the LCDM sample with resolution 3072 with
different $\pm m$ steps: for $r \le 6.5$~\Mpc\ $m =3$, for
$6.5 < r \le 17$~\Mpc\ $m=10$, and thereafter $m=50$.  In case of the
LCDM sample with resolution 1024, we used at $4.5 \le r \le 25$~\Mpc\
$m=8$, thereafter $m=25$.  The final fractal dimension function
presented in Fig.~\ref{fig:Fig3} is a combined function that we determined using
the high-resolution field 3072 at small distances, $r \le 8$~\Mpc, and
the field 1024 at larger distances.

LCDM model samples are based on all particles of the simulation and
contain detailed information on the distribution of matter in regions
of different density.  Fig.~\ref{fig:Fig1} shows
the correlation functions, $\xi(r)$,   Fig.~\ref{fig:Fig2} shows the
structure  functions,
$g(r)=1 + \xi(r)$, and  Fig.~\ref{fig:Fig3} shows  the fractal dimension functions,
$D(r)=3+\gamma(r) $ for the basic samples of the model.

The first impression from the figure is that the amplitude of the correlation
and structure functions continuously increases with increasing
particle density threshold $\rho_0$ of models.  This increase 
can be expressed by the amplitude parameter $\xi_6$, given in
Table~\ref{Tab1}.  The radius $r_z$ where the correlation function becomes
negative is equal to $r_z \approx 80$~\Mpc\ for all LCDM samples.

\begin{figure*}[ht] 
\centering 
\resizebox{0.45\textwidth}{!}{\includegraphics*{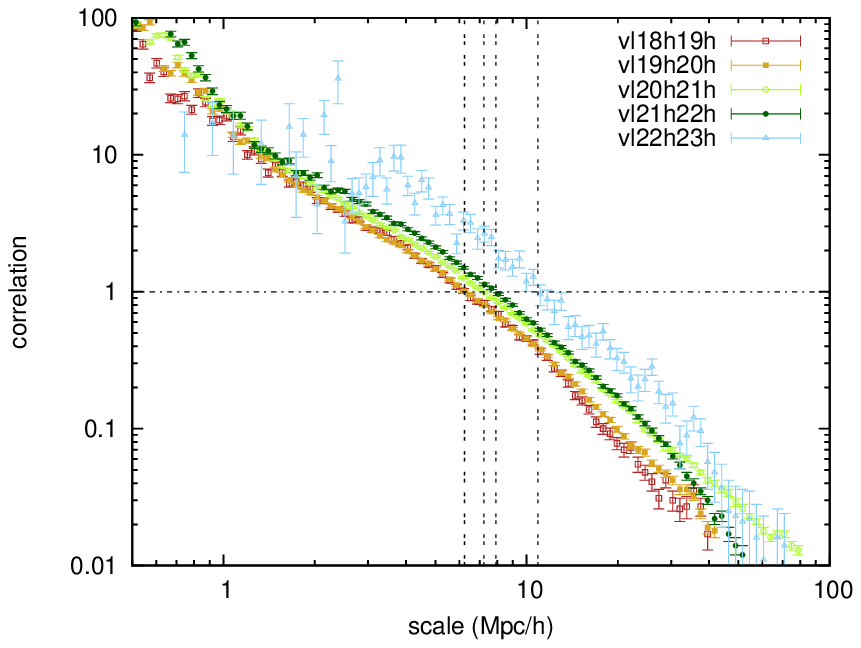}}
\resizebox{0.45\textwidth}{!}{\includegraphics*{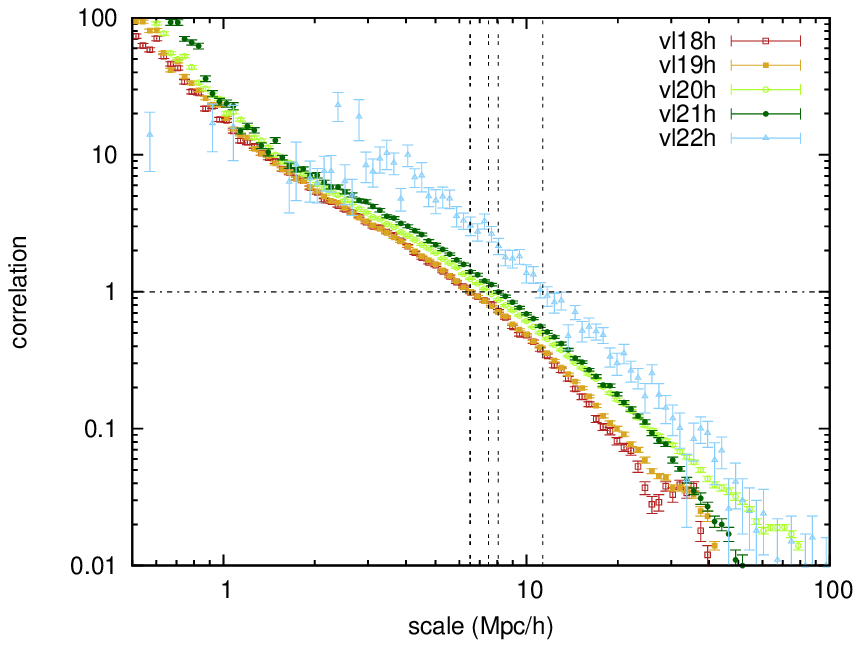}}
\caption{Comparison of correlation functions of SDSS samples found for
absolute magnitude bins and thresholds, left and right panels,
respectively. Error bars are shown.}
\label{fig:Fig4} 
\end{figure*}

{\scriptsize 
\begin{table}[ht] 
\caption{SDSS luminosity-limited samples.} 
\label{Tab2}                         
\centering
\begin{tabular}{lcrccc}
\hline  \hline
Sample   &$M_r$  & $N_{\mathrm{gal}}$ &
           $d_{\mathrm{max}}$ & $r_0$ & $\xi_6$ \\                                                           
\hline  
(1)&(2)&(3)&(4)&(5)&(6)\\ 
\hline  
SDSS.18b & $-18.0$ & 16\,676   & 135  & $6.41 \pm 0.42$ & 1.108  \\
SDSS.19b & $-19.0$ & 49\,963   & 211  & $6.31 \pm 0.24$ & 1.100 \\
SDSS.20b & $-20.0$ & 105\,088   & 323 & $7.20 \pm 0.19$ & 1.359 \\
SDSS.21b & $-21.0$ & 98\,674  & 486  & $7.93 \pm 0.21$ & 1.599  \\
SDSS.22b & $-22.0$ & 9\,430   & 571  & $11.45 \pm 0.99$ & 3.304 \\
  \\
SDSS.18t & $-18.0$ & 39\,711  & 135  & $6.55 \pm 0.28$  & 1.156  \\
SDSS.19t & $-19.0$ & 88\,132   & 211  & $6.56 \pm 0.19$  & 1.165  \\
SDSS.20t & $-20.0$ & 134\,431  & 323  & $7.46 \pm 0.17$  & 1.447 \\
SDSS.21t & $-21.0$ & 104\,512   & 486  & $8.14 \pm 0.21$  & 1.675 \\
SDSS.22t & $-22.0$ & 9\,525   & 571  & $11.32 \pm 0.97$  & 3.444  \\
\hline 
\end{tabular} 
 \tablefoot{
 The columns show the
(1) sample name; 
(2) the absolute $r-$magnitude limit, $M_r - 5\log h $;
(3) the number of galaxies in a sample;
 (4) the maximum comoving distance $d_{\mathrm{max}}$ in \Mpc;
(5) the correlation length of the sample $r_0$ in \Mpc; and
(6) the correlation function amplitude $\xi_6$ at $r_6=6$~\Mpc.
}
\end{table} 
} 

\subsection{ SDSS galaxy samples}

We used the luminosity-limited (usually referred to as volume-limited) galaxy
samples by \citet{Tempel:2014ab}, selected from data release 12 (DR12)
of the SDSS galaxy redshift survey \citep{Ahn:2014aa}.  { The authors 
complemented the  SDSS galaxy sample with redshifts from the 2MRS, 2dFGRS,
and RC3 catalogues.  The catalogue has a Petrosian $r-$band magnitude limit $m_r
\le 17.77$,  it  covers 7221 square degrees in the sky,  and contains  
584\,450 galaxies.  We used a version of the catalogue 
with a reduced footprint  in survey coordinates $-33^\circ \le \eta \le 36^\circ$,
$-48^\circ  \le \lambda  \le  51^\circ$. This version uses the most
reliable data and contains 489\,510 galaxies. For comparison, we note
that in the study of the SDSS correlation
function, \citet{Zehavi:2011aa} used a sample of 540\,000 galaxies in the magnitude interval $14.4
\le m_r \le 17.6$, in a footprint 7700\,deg$^2$.  }

The data on five luminosity-limited SDSS samples are listed in
Table~\ref{Tab2}: the limiting absolute magnitudes in red $r-$band, $M_r$,
the maximum comoving distances, $d_{\mathrm{max}}$, and the numbers of
galaxies in samples, $N_{\mathrm{gal}}$.  All SDSS samples have
a minimum comoving distance $d_{\mathrm{min}}= 60$~\Mpc.
{ At smaller distance, the catalogue is inhomogeneous, see Fig.~5 by
  \citet{Tempel:2014ab}.  The maximum comoving distances are taken from
  \citet{Tempel:2014ab}; within these limits, volume-limited samples
  have a distance-independent number density of galaxy groups.}
The SDSS samples
with $M_r - 5\log h $ luminosity limits $-18.0$, $-19.0$, $-20.0$,
$-21.0$, and $-22.0$ are referred to as SDSS.18i, SDSS.19i, SDSS.20i,
SDSS.21i, and SDSS.22i, with the additional index $i$, where $i=b$ is for
luminosity bin samples, and $i=t$ is for luminosity threshold
samples.  In luminosity bin samples, the width of the bin is one mag,
$\Delta M=1.0$, and in luminosity threshold samples, the upper limit
$M_r - 5\log h =-25.0$ exceeds the luminosity of the brightest
galaxies in the SDSS sample.  The respective luminosity limits in solar
units were calculated using the absolute magnitude of the Sun in
$r-$band, $M_\odot= 4.64$ \citep{Blanton:2007aa}.  Table \ref{Tab2}
also provides the correlation radius of samples, $r_0$, and the value
of the correlation function at $r_6=6$~\Mpc, $\xi_6$, which we used to
characterise the amplitude of the correlation function.

The SDSS galaxy samples are conical shells, and the corresponding random
particle samples were calculated using identical conical shell
configurations.  The correlation functions of SDSS galaxy samples were
calculated using the Landy-Szalay estimator.  The SDSS samples differ from model
samples in one important aspect. We used samples limited in absolute
magnitude, but observational samples are flux limited.  For this
reason, 
for the faintest galaxies, only the nearby region with a distance limit
$d_{max} = 135$~\Mpc\ can be used, in contrast to the most luminous
sample, for which the distance limit is $d_{max} = 571$~\Mpc.  The SDSS galaxy samples are
conical and have different sizes: for example, the volume of sample
SDSS.21 is about 50 times larger than the volume of sample
SDSS.18.

To determine the behaviour of the correlation function on small distances,
we used distances $0.1 \le r \le 1$~\Mpc\ using $N_B=20$ log bins,
and on scales  $0.5 \le r \le 200$~\Mpc\ using $N_B = 45$ log bins.
The fractal dimension function was calculated using the {\tt fit}
subroutine with $m=2$ for small distances, $r \le 1$~\Mpc,  and $m=4$
for larger distances. 

The correlation functions of the SDSS galaxies are shown in
Fig.~\ref{fig:Fig1} for the luminosity threshold, and in
Fig.~\ref{fig:Fig4} for the luminosity bins and thresholds. The correlation functions
for the luminosity bins are very similar to the correlation functions for
the luminosity thresholds, but the correlation lengths are slightly lower,
see Table~\ref{Tab2}.  This is expected because in threshold-limited
samples, brighter galaxies are included. These galaxies have larger correlation
lengths.

The radius $r_z$ at which the correlation function becomes negative is the same
for the SDSS.18 and SDSS.19 samples, $r_z \approx 45$~\Mpc, for brighter
galaxies, it lies higher, $r_z \approx 120$~\Mpc. This difference is
expected because the sample sizes for fainter galaxies are small and
cannot be considered as fair samples of the universe. 
The structure function $g(r)=1+\xi(r)$ continuously decreases with
increasing distance $r$.

\subsection{Millennium simulation galaxy samples}

{ We used simulated galaxy catalogues by \citet{Croton:2006aa},
which are calculated based on semi-analytical models of galaxy formation from
the Millennium simulations by \citet{Springel:2005aa}.}  The Millennium
catalogue contains 8\,964\,936 simulated galaxies, with absolute
magnitudes in {\rm r} colour $M_r \le -17.4$.  The magnitudes are in
standard SDSS filters, not in $-5\log h$ units.  We extracted the following data: $x,y,z$ coordinates, and
absolute magnitudes in $r$ and $g$ photometric systems.  The Millennium
samples with $M_r $ luminosity limits
$-18.0,~-19.0,~-20.0,~-20.5,~-21.0, ~-21.5$, $-22.0$ and $-23.0$ are
referred to as Mill.18i, Mill.19i, Mill.20i, Mill.20.5i, Mill.21i,
Mill.21.5i, Mill.22i, and Mill.23i, where the index $i=b$ is for
luminosity bin samples, and $i=t$ is for luminosity threshold samples
with an upper limit $-25.0$.  Data on the Millennium luminosity-limited
samples are given in Table~\ref{Tab3}, including the correlation lengths
found for Millennium galaxy samples, and the value of the correlation
function at $r_6=6$~\Mpc, $\xi_6$.

{\scriptsize 
\begin{table}[ht] 
\caption{Millennium luminosity-limited galaxy samples.} 
\label{Tab3}                         
\centering
\begin{tabular}{lcrcc}
\hline  \hline
Sample   & $M_r $  & $N_{\mathrm{gal}}$ &
           $r_0$ & $\xi_6$ \\                                                           
\hline  
(1)&(2)&(3)&(4)&(5)\\ 
\hline  
Mill.18b & $-18.0$ &  829\,725   & $5.45 \pm 0.05$  & 0.907  \\
Mill.19b & $-19.0$ & 1\,022\,303     & $5.70 \pm 0.03$  & 0.954 \\
Mill.20b & $-20.0$ &
                                 1\,209\,900  & $5.60 \pm 0.04$  & 0.955  \\
Mill.21b & $-21.0$ &  568\,705  & $5.44 \pm 0.06$  & 0.902 \\
Mill.22b & $-22.0$ & 98\,575     & $6.54 \pm 0.18$  &  1.176\\
Mill.23b & $-23.0$ &  5\,414   & $12.8 \pm 1.46$  &   3.574 \\
  \\
Mill.20.0t & $-20.0$ & 1\,882\,809   & $5.79 \pm 0.04$  & 0.952  \\
Mill.20.5t & $-20.5$ & 1\,183\,825   & $5.81 \pm 0.04$  & 0.956 \\
Mill.21.0t & $-21.0$ &  672\,909  & $5.81 \pm 0.06$  & 0.956  \\
Mill.21.5t & $-21.5$ & 309\,580   & $6.13 \pm 0.09$  & 1.033  \\
Mill.22.0t & $-22.0$ &  104\,204   & $6.84 \pm 0.18$  & 1.271  \\
Mill.23.0t & $-23.0$ &  5\,629   & $12.9 \pm 1.44$ & 3.766 \\
\hline 
\end{tabular} 
 \tablefoot{
 The columns show the
(1) sample name; 
(2) the absolute $r-$magnitude limit, $M_r $;
(3) the number of galaxies in a sample;
(4) the correlation length of the sample $r_0$ in \Mpc; and
(5)  the correlation function amplitude at $\xi_6$. 
}
\end{table} 
} 

The Millennium sample of simulated galaxies has almost nine million
objects.  Our first task was to calculate correlation functions for
such a large dataset.  We tried several possibilities: dividing the
sample into smaller subsamples of 100 and 250~\Mpc, and using
smaller luminosity bins.  Our experience shows that boxes of 250 and
500~\Mpc\ yield almost identical results for the 
correlation function. The sample of 100~\Mpc\ contains too
few galaxies, which leads to larger random and some systematic errors.

One possibility to reduce the size of galaxy samples is to use
smaller magnitude bins.  At the low end of the luminosity
range, we tried bins of  0.15, 0.25, and 0.5 in absolute magnitude.  The results
for the correlation function were practically identical.  For
samples Mill.18b and Mill.19b, we used luminosity bins 0.25 and 0.50 mag,
respectively, and for the brighter samples, we used bins of 1.0 mag. The samples with threshold
magnitudes at low absolute magnitudes are so large that our program
for computing the correlation function was not able to manage the
file. For this 
reason, we only calculated the correlation functions for magnitude thresholds
for samples with $M_r \le -20.0$.  The results are listed in
Table~\ref{Tab3}.  In Fig.~\ref{fig:Fig1},  Fig.~\ref{fig:Fig2}, and
Fig.~\ref{fig:Fig3} we show the correlation 
function, $\xi(r)$, the structure function, $1+\xi(r)$, and the
fractal dimension function, $D(r)=3+\gamma(r)$ for the luminosity bins.  The
number of simulated galaxies in most  magnitude bin samples is so
large that the correlation functions are very smooth with almost no
scatter, thus it is easy to find the correlation length $r_0$ and the
amplitude parameter $\xi_6$ of samples by linear interpolation. As
usual, the correlation length is defined as the distance at which the
correlation function has the unit value,$\xi(r_0)=1.0$.  At this
distance, the central galaxy has just one neighbour, and the density of
galaxies is twice the mean density.

We calculated the correlation functions with three
resolutions at distances $0.1 \le r \le 10$~\Mpc,
$1 \le r \le 100$~\Mpc, and $2 \le r \le 200$~\Mpc, all with
$N_b = 100$ equal linear bins of 0.1,~1, and 2~\Mpc,
respectively.  The correlation functions presented in Fig.~\ref{fig:Fig1}
are the results, combined from resolutions with bin sizes 0.1 and 1~\Mpc,
used for $r \le 3$ and $r >3$~\Mpc, respectively.  To determine the fractal
dimension function, we applied smoothing with  $m=3$ up to
$r = 10$~\Mpc, and $m=6$ on larger distances.

\begin{figure}[ht] 
\centering 
\hspace{1mm}  
\resizebox{0.45\textwidth}{!}{\includegraphics*{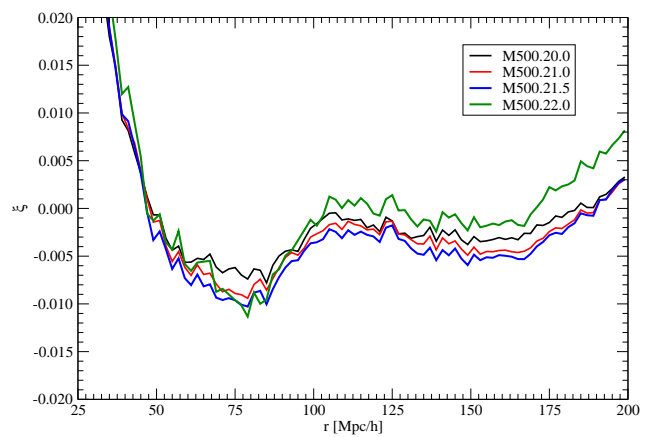}}
\caption{ Correlation functions on large scales of Millennium simulations  for luminosity
  thresholds to show the BAO peak at $r \approx 110$~\Mpc. }
\label{fig:Fig5} 
\end{figure} 

Our calculation show that the correlation function of Millennium
galaxies at small and medium distances has a continuously changing
gradient  $\gamma(r)$, as found for the SDSS samples.  At larger distances, the
correlation function bends and crosses the zero value, $\xi(r_z)=0$.
The radius $r_z$ where the correlation function becomes negative lies at
$r_z = 46 \pm 1$~\Mpc\ for all samples, see Fig.~\ref{fig:Fig1}.

On large scales $r \ge r_z$ , the correlation function lies close to
the zero value. However, the baryonic acoustic oscillation (BAO) peak at
$r \approx 110$~\Mpc\ is clearly visible when the function is 
 expanded in vertical direction, as shown in 
 Fig.~\ref{fig:Fig5}. 

\subsection{EAGLE simulation galaxy samples}

The Evolution and Assembly of GaLaxies and their Environments (EAGLE) is a
suite of cosmological hydrodynamical simulations that follow the
formation and evolution of galaxies. Its goals are described by
\citet{Schaye:2015aa} and \citet{Crain:2015aa}.  We extracted galaxy
data from the public release of halo and galaxy catalogues
\citep{McAlpine:2016aa}.  The EAGLE simulations were run with the following
cosmological parameters: $\Omega_m =0.308$, $\Omega_{\Lambda}=0.692$, and
$H_0 = 67.8$ km/s per Megaparsec. Within the EAGLE project, several
simulations were run. We used for the correlation analysis the
simulation within a box of comoving size $L=100$~Mpc.  In contrast to
convention, box sizes and lengths are not quoted in units 
of $1/h$.  This means that in conventional units, the size of the
simulation box in comoving coordinates is $L=67.8$ ~\Mpc.  We used
$x,y,z$ coordinates and absolute magnitudes 
in the $r$ and $g$ photometric systems.

{\scriptsize 
\begin{table}[ht] 
\caption{EAGLE luminosity-limited samples.} 
\label{Tab4}                         
\centering
\begin{tabular}{lcrcc}
\hline  \hline
Sample   &$M_r$  & $N_{\mathrm{gal}}$ &
           $r_0$ & $\xi_6$ \\                                                           
\hline  
(1)&(2)&(3)&(4)&(5)\\ 
\hline  
EAGLE.16b & $-16.0$ & 7\,887   & $4.72 \pm 0.45$  &  0.628\\
EAGLE.17b & $-17.0$ & 5\,553   &  $4.81 \pm 0.54$  & 0.665 \\
EAGLE.18b & $-18.0$ & 4\,088    & $4.72 \pm 0.62$  &  0.684 \\
EAGLE.19b & $-19.0$ & 3\,025    & $5.15 \pm 0.79$ & 0.769  \\
EAGLE.20b & $-20.0$ & 2\,122  & $5.35 \pm 0.98$& 0.847\\
EAGLE.21b & $-21.0$ & 987   & $5.02 \pm 1.34$ & 0.789  \\
EAGLE.22b & $-22.0$ & 204    & $6.45 \pm 3.79$ & 0.868  \\
  \\
EAGLE.16t & $-16.0$ & 23\,881   &  $4.86 \pm 0.26$ & 0.703 \\
EAGLE.17t & $-17.0$ & 15\,994  & $4.95 \pm 0.33$ & 0.736  \\
EAGLE.18t & $-18.0$ & 10\,441   & $5.04 \pm 0.41$ & 0.772  \\
EAGLE.19t & $-19.0$ & 6\,353   & $5.36 \pm 0.56$ & 0.837 \\
EAGLE.20t & $-20.0$ & 3\,328   & $5.36 \pm 0.78$ & 0.891  \\
EAGLE.21t & $-21.0$ & 1\,206   &  $5.85 \pm 1.42$ & 0.907 \\
EAGLE.22t & $-22.0$ &  219     & $7.46 \pm 4.23$ &  0.868  \\
\hline 
\end{tabular} 
 \tablefoot{
 The columns list
(1) the sample name; 
(2) the absolute $r-$magnitude limit, $M_r  $;
(3) the number of galaxies in a sample;
(4) the correlation length of the sample $r_0$ in \Mpc; and
(5) the correlation function amplitude at $\xi_6$.
}
\end{table} 
}

The EAGLE samples with $M_r  $ luminosity limits
$-15.0,~-16.0,~-17.0,~-18.0,~-19.0,~-20.0, ~-21.0$, and $-22.0$ are referred
to as EAGLE.15i, EAGLE.16i, EAGLE.17i, EAGLE.18i, EAGLE.19i, EAGLE.20i, EAGLE.21i,
and EAGLE.22i, where the index $i=b$ is for luminosity bin
samples of a bin size of 1.0 mag, and $i=t$ is for luminosity threshold samples with an upper
limit  of $M_r - 5\log h = -25.0$.  Data on EAGLE luminosity-limited samples are given in
Table~\ref{Tab4}. We also list the correlation lengths found for
EAGLE galaxy samples, and the value
of the correlation function at $r_6=6$~\Mpc, $\xi_6$.

For the correlation analysis we used the largest simulated galaxy
sample with 29\,730 objects in the $M_r$ luminosity interval
$-14.7 \ge M_r \ge -23.7$.  This allowed us to calculate correlation
functions for luminosity bins and luminosity threshold limits starting
from $M_r = -15.0$ up to $M_r = -22.0$.  The data of the samples are listed in
Table~\ref{Tab4}. The correlation, structure, and fractal dimension
functions are presented in Figs.~\ref{fig:Fig1}  --  \ref{fig:Fig3}  for luminosity
thresholds.  The distance scale is expressed in \Mpc.

The correlation functions of the EAGLE samples were calculated for two
resolutions: at distances $0.2 \le r \le 20$~Mpc and
$1 \le r \le 100$~Mpc, with $N_b = 100$ equal linear bins of size 0.2
and 1~Mpc, respectively.  The correlation functions are presented in
Fig.~\ref{fig:Fig1}.  High-resolution functions were used up to
distance $r \le 3$~\Mpc, and low-resolution functions on larger
distances. To determine the fractal dimension function, we applied a smoothing
with $m=3$ up to $r = 3$~\Mpc, and $m=15$ on larger distances.

The shape of the correlation functions is similar to that of those for the
Millennium simulation.  The basic difference is that for the EAGLE
simulation, it was possible to derive correlation functions for much
fainter samples up to $M_r=-15.0$.  The radius $r_z$ where the correlation
function becomes negative is the same for all samples
$r_z = 17 \pm 1$~\Mpc. This is much smaller than for the SDSS, LCDM, and Millennium
samples. This is due to the small volume of the simulation box.

\subsection{Error estimations of the correlation functions}

The sampling errors we found using the bootstrap method for the SDSS samples are shown
in Fig.~\ref{fig:Fig4} for the absolute magnitude bin and the threshold
samples in the left and right panels. From these data we also determined the errors
of the correlation length, $r_0$, given in Table~\ref{Tab2}.  We found
that the sampling errors can approximately be expressed as follows:
$\epsilon(r_0) = a \times r_0 /\sqrt{N}$, where $N$ is the number of
galaxies in the sample, and $a$ is a constant.  The correlation
lengths of SDSS galaxies behave similarly, as found by 
\citet{Zehavi:2011aa}.  A similar relation can also be used for the errors
of the correlation function amplitudes, $\xi_6$.  Using the 
\citet{Zehavi:2011aa} SDSS sample and our sample, we found for the parameter the value
$a=8.4$.  Using this relation, we estimated errors of the correlation
lengths of the Millennium and EAGLE simulation samples given in Tables.
  The relative errors of the correlation amplitudes
are equal to the relative errors of the correlation lengths.

Most of our samples contain so many galaxies that the random sampling errors
are very small, and the correlation and structure functions are very
smooth.  For this reason, the scatter of the correlation lengths, $r_0$,
and correlation amplitudes, $\xi_6$, that we found for various luminosity
limits of identical samples is also small, see Tables \ref{Tab1} to
\ref{Tab4}.  The relative  sampling errors of the  correlation length, $r_0$,
and the correlation amplitude, $\xi_6$, are of the order of a few
percent for most samples.  Samples corresponding to luminous galaxies 
contain fewer objects, therefore the random sampling errors of $r_0$ and $\xi_6$ are
larger.

The number of galaxy-galaxy pairs at large distances is very large, as
is the number of random-random pairs.  Correlation functions were
calculated based on the ratios of very large almost identical numbers.  For
this reason, the sampling errors of the correlation functions on large
distances are larger, and larger smoothing was needed to calculate
the correlation dimension functions.

\section{Biasing properties of LCDM and galaxy samples}

In this section we give an overview of previous determinations of
the correlation functions. Next we discuss the luminosity dependence of
the correlation functions and their derivatives of LCDM and galaxy samples,
and the bias parameters that were found for these samples.   We
focus the discussion on the differences of the correlation properties in
samples with real or simulated galaxies and with only DM particles.

\subsection{Correlation functions of galaxies: previous studies}

Classical methods for calculating angular and spatial correlation
functions using catalogues of extragalactic objects were elaborated by
\citet{Limber:1953uq, Limber:1954fj}, \citet{Totsuji:1969aa},
\citet{Peebles:1973a}, and \citet{Groth:1977}, see also
\citet{Peebles:1980aa} and \citet{Martinez:2002fu}.  An analysis of
almost all galaxy and cluster catalogues that were available in the 1970s
(\citet{Abell:1958}, \citet{Shane:1967}, and \citet{Zwicky:1968}) was
performed by  \citet{Hauser:1973}, \citet{Peebles:1974aa}, \citet{Peebles:1975},
  \citet{Peebles:1975a}, \citet{Peebles:1975aa},  \citet{Groth:1977},  and \citet{Seldner:1977aa}.

{ These studies showed that the spatial
correlation function, calculated from the angular correlation function,
can be expressed by a simple law: 
\be
\xi(r)=(r/r_0)^\gamma,
\label{xilaw}
\ee
where $r_0 = 4.7$~\Mpc\ is the
correlation length of the sample, and the power index is
$\gamma = -1.77$ for the Shane-Wirtanen catalogue \citep{Groth:1977}. 
Other catalogues studied  yield almost the same parameters.

In the 1970s, redshift surveys of galaxies were complete enough to
determine the spatial correlation functions of galaxies and galaxy
clusters. 
To use redshift data, \citet{Peebles:1976aa}, \citet{Davis:1978aa}, and 
  \citet{Davis:1983ly} suggested to use the galaxy position and velocity
information separately.  In this case, pair separations can be
calculated parallel to the line of sight, $\pi$, and perpendicular to
the line of sight, 
$r_p$. The angular correlation function, $w_p(r_p)$,
can be found by integrating over the measured $\xi(r_p,\pi)$, using
the equation
\begin{equation}
  w_p(r_p) = 2 \int_{r_{min}}^{r_{max}} \xi(r_p, \pi) d\pi,
  \label{wp}
\end{equation}
where $r_{min}$ and $r_{max}$ are minimum and
maximum distances of the galaxies in samples.  
This equation has the
form of the Abel integral equation, and can be inverted to recover the
spatial correlation function \citep{Davis:1983ly},
\begin{equation}
 \xi(r) = - {1 \over \pi} \int_r^{r_{max}} {w_p(r_p) \over \sqrt{r_p^2 -
     r^2}} d r_p.
 \label{xi}
\end{equation}
\citet{Davis:1983ly} found that the spatial correlation
function can be well represented by a power law Eq.~(\ref{xilaw}), with
the parameters $r_0=5.4 \pm 0.3$~\Mpc, and $\gamma = -1.77$.

\citet{Klypin:1983fk} and \citet{Bahcall:1983uq} found that the
correlation function of rich clusters of galaxies is similar to the
correlation function of galaxies, but has a much larger correlation
length.  This behaviour has been predicted by \citet{Peebles:1974aa}.
This difference was explained by \citet{Kaiser:1984} as the
biasing phenomenon: rich clusters are more clustered or more biased
representatives of extragalactic objects.

\citet{Einasto:1986ab} studied the effect of voids on spatial
correlation function. The authors showed that the presence of voids
increases the correlation amplitude and length. This is clear from a
comparison of two simple toy models: the galaxies in the models are identical, but
in the void model, the sample is surrounded by an empty space. In
the void sample, the same number of random points fills a larger volume,
and in a given separation interval, there are fewer random points than
in the first sample. This increases the correlation amplitude
according to Eq.~(\ref{nat}).  \citet{Einasto:1986ab} argued that Lick
counts represent the numbers of galaxies integrated over the line of sight
in the Lick maps, but voids that are present in the volume along the line of
sight are filled with foreground and background objects. This causes
the decrease in amplitude of the correlation functions found in 
2D data. The authors also showed that the correlation length increases
for more  luminous  galaxies.

These results were interpreted by \citet{Pietronero:1987aa} and
\citet{Calzetti:1987aa,Calzetti:1988aa} as the presence of fractal
structure in the universe.  \citet{Hamilton:1988aa} confirmed the
dependence of the correlation amplitude on the  luminosity of
galaxies.  The  amplitude of the correlation functions of galaxies
brighter than $M_B$ relative to the amplitude at $M_B = -19$ is
almost flat for $M_B > -20$ and rises considerably for $M_B <
-21$. He explained this as the biasing effect: luminous galaxies form
in more clustered regions.

\citet{Maddox:1990aa, Maddox:1996aa} measured the angular correlation
function of the APM Galaxy Survey, which contains over two million
galaxies in the magnitude range $17 \le b_j \le 20.5$ in the
photometric system $b_j$ of the survey. At small separations, the
angular correlation function was determined by counting individual
galaxies, and on larger separations by counts of galaxies in cells of size
$0.^\circ23 \times 0.^\circ23$.  Estimates of the angular correlation
function $w(\theta)$ were made in six magnitude slices.  The data
agree well with a power law $w(\theta) = A \theta^{\gamma-1}$, 
where $A$ is a constant depending on sample depth. The authors found that this
power law is valid over a wide range of angular separations $0.01 \le
\theta \le 3$ degrees, with a constant slope $\gamma = -1.699 \pm
0.032$, in good agreement with \citet{Groth:1977} for the Lick survey. The
amplitude of the  correlation function was calculated using the Limber
equation and applying  cosmological models with various parameters.
The resulting correlation length varies between $r_0 = 4.4 - 5.0$~\Mpc,
depending on the model. 

\citet{Norberg:2001aa} and \citet{Hawkins:2003aa} used the 2dF Galaxy
Redshift Survey (2dFGRS) to calculate the correlation functions of
galaxies and to investigate the clustering properties of the
universe. The authors analysed in detail the two-point correlation
function $\xi(r_p,\pi)$ and its spherical average, which gives an
estimate of the redshift-space correlation function, $\xi(s)$, where
$s$ is the separation of galaxies in redshift space. As the first step, the
authors measured the 2D correlation function $\xi(r_p,\pi)$, and
integrated over the line of sight using Eq.~(\ref{wp}) to avoid
supercluster contraction effects \citep{Kaiser:1987aa}.  The authors found
that the effective
depth of the 2dFGRS is $z_s \approx 0.15$, the effective luminosity
$L_s \approx 1.4L_\star$, the redshift-space clustering length is $s_0=6.82
\pm 0.28$~\Mpc, and the real-space correlation length is $r_0=5.05 \pm
0.26$~\Mpc. \citet{Norberg:2001aa} found for the relative bias
function of galaxies $b/b_\ast = 0.85 + 0.15 L/L_\star$, which agrees well with \citet{Hamilton:1988aa}.

\citet{Zehavi:2004aa, Zehavi:2005aa, Zehavi:2011aa} studied the luminosity dependence
of the correlation functions using flux-limited and volume-limited samples
of SDSS galaxies.  The authors measured the projected 2D correlation function
$w_p(r_p)$.  In the first step, the authors calculated $\xi(r_p,\pi)$ by
counting data-data, data-random, and random-random pairs using the
Landy-Szalay estimator Eq.~(\ref{LS}). Next, the authors computed the
projected correlation functions $w_p(r_p)$ using Eq.~(\ref{wp}), and
adopting $r_{min}=0$ and $r_{max}=40$~\Mpc.  \citet{Zehavi:2011aa}
found for the large-scale bias function the form
$b_g(>L) \times (\sigma_8/0.8) = 1.06 + 0.21(L/L_\ast)^{1.12}$, where
$L$ is the $r-$band luminosity corrected to $z=0.1$, and $L_\ast$
corresponds to $M_\ast=-20.44 \pm 0.01$ \citep{Blanton:2003aa}.

\citet{Zehavi:2004aa, Zehavi:2005aa, Zehavi:2011aa} concentrated
their effort on investigating the profile of the correlation function. The authors
used for comparison a $\Lambda$CDM model by applying the halo occupation
distribution (HOD) model by \cite{Peacock:2000aa},
\citet{Seljak:2000aa}, and \citet{Berlind:2002aa}. The authors assumed that
halos have a Navarro-Frenk-White (NFW) profile \citep{Navarro:1996aa}, and are populated with
galaxies that trace the DM profile. The authors found that correlation
functions of HOD models and luminosity-limited SDSS samples
change in the slope of the correlation function at $r \approx
2$~\Mpc. On a smaller scale, the one-halo contribution dominates, and on
a larger scale, the galaxies from different halos dominate.  We note that
this effect has been described by \citet{Zeldovich:1982kl} and
by \citet{Einasto:1991oy, Einasto:1992aa} using galaxy
data. \citet{Einasto:1992aa} found that the transition from clusters
to filaments with different values of the slope of the correlation
function is rather sharp: $\gamma=-1.8$ for $r < 3$~\Mpc, and
$\gamma=-0.8$ for larger scales. This sharp transition was confirmed
by \citet{Zehavi:2004aa} on the basis of the  HOD model.  }

\begin{figure}[ht] 
\centering 
\hspace{1mm}  
\resizebox{0.46\textwidth}{!}{\includegraphics*{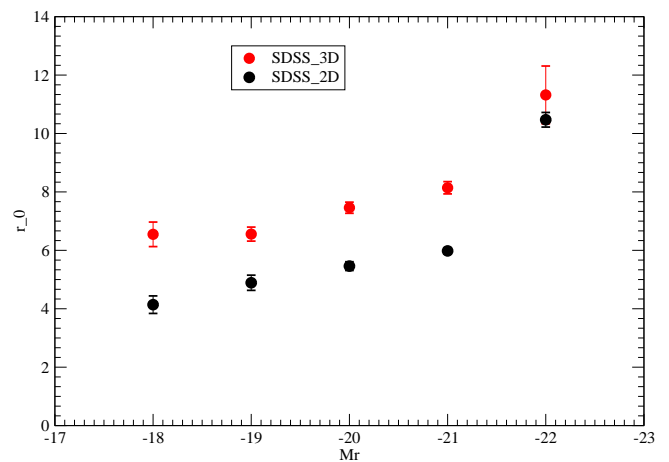}}
\caption{Correlation length $r_0$ of SDSS galaxies according to 2D
  data by \citet{Zehavi:2011aa} and 3D data in this paper, using
  luminosity  thresholds samples. }
\label{fig:Fig6a} 
\end{figure} 

{ We compare in Fig.~\ref{fig:Fig6a} the correlation lengths of SDSS
  samples of various luminosity.  \citet{Zehavi:2011aa} found spatial
  correlation functions from projected correlation functions as
  described above.  Our correlation functions were calculated directly
  from 3D data.  Here two trends are clearly visible:  the
  dependence of the correlation lengths on the galaxy luminosity, and
  the overall shift of the amplitudes of 3D correlation functions with
  respect to the correlation functions from a 2D analysis. The luminosity
  dependence of 2D and 3D samples is similar.  The general shift of
  the amplitudes of the 3D correlation functions with respect to 2D
  samples arises because in 3D samples, the regions of zero spatial
  density 
  are taken into account and increase the amplitude of the correlations
  functions.  In a 2D analysis, the presence of voids is
  ignored. The correlation lengths of SDSS galaxies in identical
  luminosity limits are according to our 3D analysis in the mean 1.35
  times higher than according to the 2D analysis by
  \citet{Zehavi:2011aa}.}

\subsection{Luminosity dependence of the correlation functions of the SDSS,
  Millennium, and EAGLE samples}

The luminosity dependence of the correlation functions is the principal
factor of the biasing phenomenon, as discussed by \citet{Kaiser:1984}.
Here we compare the luminosity dependence of the
correlation functions of real and simulated galaxies, as found in the
present analysis.

Our first task was to compare the clustering properties of real SDSS galaxy
samples with clustering properties of simulated galaxies of the EAGLE and
Millennium simulations.  { As noted above, in simulations, the absolute
  galaxy magnitudes were not computed in $-5\log h$ units. To
  compare the correlation functions of 
  simulated and real samples, the simulated galaxy data must be reduced to
  real galaxy data in terms of luminosities.  The simulated galaxy samples
  of the EAGLE and Millennium models for various luminosity limits have
  spatial densities that differ from the spatial densities of SDSS
  samples for these luminosity limits. To bring model galaxy samples
  to the same system as SDSS samples, the luminosities of the simulated
  galaxies must be weighted on the basis of the comoving luminosity
  densities of samples. A similar procedure was applied by
  \citet{Liivamagi:2012aa} to reduce the luminosity scale of luminous
  red giants (LRG) to the main sample of SDSS galaxies.  We found
comoving spatial densities of galaxies for all SDSS, EAGLE, and
Millennium subsamples for various luminosity limits.  The comparison
of densities shows that in order to bring spatial densities of
simulated galaxies to the same density level as that of the SDSS galaxies,
the luminosities  of the weighted EAGLE simulation samples must be
decreased by shifting the absolute magnitudes   toward fainter magnitudes by
0.38 mag.  The decrease in magnitudes of  the Millennium 
simulation samples is  0.80 mag.  To obtain identical spatial densities,
the shift is slightly luminosity  dependent, but we ignored this
difference and used identical shifts for all samples of galaxies of the
same simulation.

The correlation lengths for luminosity bin samples were calculated for
bins of 1 mag. The correlation lengths listed in Table~\ref{Tab2}
correspond to galaxies that are 0.5 mag brighter than given in column (2), thus
the correlation length of $L_\star$ galaxies is practically equal to
the correlation length of the sample SDSS.20b with mean magnitude,
$M_r=-20.5$, very close to the magnitude of $L_\star$ galaxies,
$M_\ast=-20.44$. Thus we accept using data from Table~\ref{Tab2},
$r_0(L_\star) = 7.20 \pm 0.19$.  We note that this value is rather
close to the value of the mean redshift space correlation length for
2dFGRS galaxies, $s_0=6.83 \pm 0.28$ \citep{Hawkins:2003aa}. }

\begin{figure}[ht] 
\centering 
\hspace{1mm}  
\resizebox{0.45\textwidth}{!}{\includegraphics*{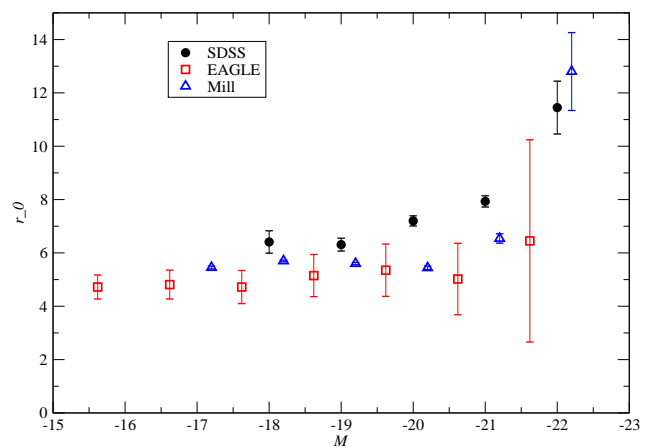}}
\caption{Correlation length $r_0$ of SDSS galaxies. For comparison, we
  also show the correlation lengths of the EAGLE and Millennium simulations for the luminosity bins. }
\label{fig:Fig6} 
\end{figure}

In Fig.~\ref{fig:Fig6} we show the correlation lengths, $r_0$, of the SDSS,
EAGLE, and Millennium samples as functions of magnitude $M_r$.  The
correlation length $r_0$ has a rather similar luminosity 
dependence for all samples.  The correlation lengths and amplitudes of
the SDSS samples are slightly higher than in the simulated galaxy samples, by a
factor of about 1.2.

An important detail in the luminosity dependence is the fact that
the correlation lengths and amplitudes are almost constant at low
luminosities, $M \ge -20.0$.  This tendency was noted earlier by
\citet{Einasto:1986ab}, \citet{Hamilton:1988aa},
\citet{Norberg:2001aa}, and \citet{Zehavi:2005aa,Zehavi:2011aa}.  It
is seen in observational samples SDSS.19 and SDSS.18, but cannot
follow to fainter luminosities because very faint
galaxies are absent from SDSS luminosity-limited samples.  In the EAGLE and Millennium
model galaxy samples, this very slow decrease of $r_0$ and $\xi_6$ with
decreasing luminosity can be followed until very faint galaxies,
$M \approx -15.6$ for the EAGLE sample, and $M \approx -17.2$ for
the Millennium sample.  This tendency indicates that very faint galaxies
follow the distribution of brighter ones, that is,  faint galaxies are
satellites of bright galaxies.  The tendency of the clustering dwarf
galaxies to form satellites of brighter galaxies has been known for a long time
\citep{Einasto:1974b}, see also \citet{Wechsler:2018fj}.  The
correlation study confirms this tendency.

 \begin{figure}[ht] 
\centering 
\hspace{1mm}  
\resizebox{0.45\textwidth}{!}{\includegraphics*{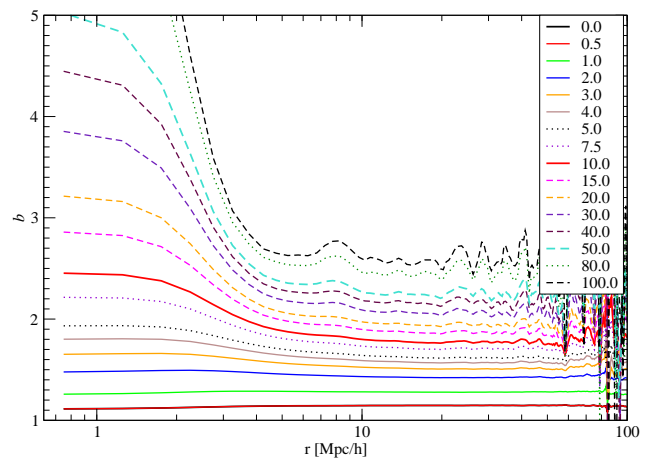}}
\caption{Bias functions, $b(r)$,  of biased LCDM models as functions of the
  galaxy pair separation $r$. Particle density limits $\rho_0$ are
  indicated as symbol labels. }
\label{fig:Fig7} 
\end{figure}

\subsection{Bias parameters of   the LCDM  and  SDSS samples}

{ In Fig.~\ref{fig:Fig7} we present  the bias functions (relative
  correlation functions) $b(r,\rho_0)$ for a series of LCDM models
  with various particle density limits $\rho_0$.}  They were calculated
using Eq.~(\ref{bias}) for the whole range of distances $r$
(galaxy separations) by dividing the correlation functions of the biased
models LCDM.$i$ by the correlation function of the unbiased model
LCDM.00.  The bias functions for the same biased LCDM models were
calculated from power spectra by \citet{Einasto:2019aa}, see Fig.~2 of
this paper.  Fig.~\ref{fig:Fig7} shows that  the bias function has a
plateau at $r_6 \approx 6$~\Mpc, similar to the plateau around
$k \approx 0.03$~$h$~Mpc$^{-1}$ of the relative power spectra
\citep{Einasto:2019aa}.  Both plateaus were used to determine the amplitudes of
the correlation functions or power spectra to define the respective bias
parameters.  The figure shows that the amplitude of the correlation
functions at small distances (pair separations) rises rapidly with the
increase in particle density limit $\rho_0$.

Fig.~\ref{fig:Fig7} shows that  bias function, $b(r)$,
for low values of the particle density threshold, $\rho_0 \le 3$, is
almost constant.  This suggests that  the correlation functions of
faint galaxies are almost parallel to the correlation function of
matter.  The only difference is in the amplitude.  At
these low values of the particle density threshold $\rho_0$ , the bias
parameter is essentially determined by the fraction of matter in the
clustered population, that is,\  matter, associated with galaxies.  This
property allows us to derive absolute bias parameters of the biased model
samples with respect to matter, as given in Table~\ref{Tab1}.  As
expected, the bias parameters found for the correlation functions are very
close to the bias parameters of the same model that were derived from power spectra
by \citet{Einasto:2019aa}.  For a more detailed discussion of the
effect of low-density regions of the density field of matter on the
correlation function and power spectrum, see
\citet{Einasto:1993ab, Einasto:1994aa, Einasto:1999aa}.

\begin{figure}[ht] 
\centering 
\hspace{1mm}  
\resizebox{0.45\textwidth}{!}{\includegraphics*{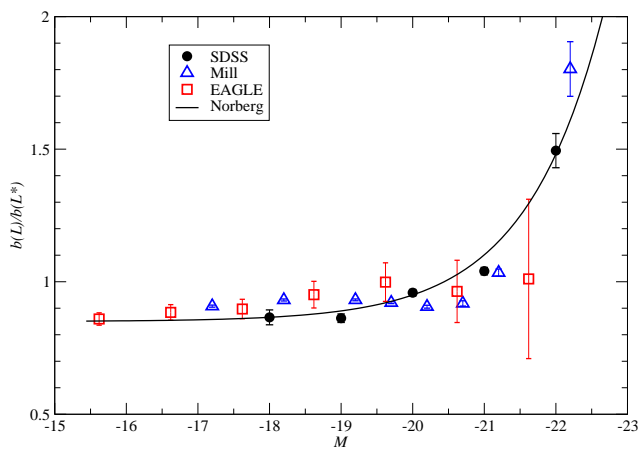}}
\caption{Relative bias function $b(L)/b(L_\ast)$ for the SDSS, EAGLE, and
  Millennium samples.  The solid line shows the relative bias function by
  \citet{Norberg:2001aa}.}
\label{fig:Fig8} 
\end{figure}

In Fig.~\ref{fig:Fig8} we show the relative bias function
$b(L)/b(L_\ast)$ for galaxies of various luminosity of the SDSS, EAGLE, and
Millennium samples. The relative bias function was
calculated from the relation
\begin{equation}
  b(L)/b(L_\ast) = \sqrt{\xi(6,L)/\xi(6,L_\ast)}\end{equation}
using amplitudes of the correlation functions at the plateau at
$r_6=6$~\Mpc\ of samples of luminosity $L$, divided by the
amplitudes of the correlation functions at  $r_6=6$~\Mpc\    of 
the same sample at luminosity $L_\ast$.  This representation is
commonly used, here $L_\ast$ is the characteristic luminosity of
the samples. It corresponds approximately to the red magnitude $M_r =
-20.5$.  The relative bias function  rises rapidly with
the increase in luminosity over $L_\ast$.  This increase in the
relative bias function  at high luminosities has been observed by
\citet{Einasto:1986ab}, \citet{Hamilton:1988aa}, 
\citet{Norberg:2001aa}, \citet{Tegmark:2004aa}, and
\citet{Zehavi:2005aa, Zehavi:2011aa}.  Fig.~\ref{fig:Fig8} shows that
within the errors, the relative bias function of the SDSS sample
agrees with the relative bias function by \citet{Norberg:2001aa}. 
Deviations of the Millennium and EAGLE relative correlation functions from
the Norberg fit are slightly larger, but within reasonable limits.

Table~\ref{Tab1} shows the bias parameter $b(\rho_0)$ of the LCDM
model as a function of $\rho_0$.  This bias parameter relates biased
models with unbiased models, that is, relative to all matter. The
bias parameter increases continuously with the increase in
threshold $\rho_0$, in contrast to the relative bias parameters of
the observational and the model EAGLE and Millennium galaxy samples,
that is, the bias parameters of $L$ galaxies with respect to $L_\ast$ galaxies,
which have flat $r_0(M)$ and $\xi_6(M)$ curves at low luminosities.

This difference can be explained by differences in the distribution of
DM particles and dwarf galaxies.  Removing low-density particles from the
LCDM sample with increasing $\rho_0$ limit affects the
distribution of high-density regions (clusters in terms of the
percolation analysis by \citet{Einasto:2018aa}) over the whole volume of
the LCDM model.  Removing very faint dwarf galaxies form
real SDSS or simulated EAGLE and Millennium galaxy samples
only affects the structure of halos that contain brighter galaxies, but not the
structure of voids because voids do not contain isolated dwarf galaxies. This
difference is rather small and does not affect the overall
geometrical properties of the cosmic web of real and model samples, as
measured by the extended percolation analysis by \citet
{Einasto:2019aa}.  However, the correlation function is sensitive
enough to detect this small difference in the spatial distribution of
dwarf galaxies and DM particles in low-density regions.

\subsection{Correlation function amplitudes of the LCDM and SDSS
  samples}

The  amplitudes of the correlation functions of the biased LCDM samples are systematically
higher than respective amplitudes of the SDSS galaxy samples, see Tables
\ref{Tab1} and \ref{Tab2} and Fig.~\ref{fig:Fig6}.
A comparison of the percolation properties of SDSS and LCDM samples showed
that LCDM models with a particle density threshold limits
$3 \le \rho_0 \le 10$ correspond approximately to SDSS samples with
absolute magnitude limits $-18.0 \ge M_r \ge -21.0$
\citep{Einasto:2018aa}.  Within these selection parameter limits, the
correlation amplitudes of biased LCDM samples can be reduced to the
amplitude levels of SDSS samples using an amplitude reduction factor
0.65.  The reason for this difference is not clear. One possibility is
that our LCDM model has a different $\sigma_8$ normalisation than real
SDSS data.

\begin{figure}[h] 
\centering 
\hspace{1mm}  
\resizebox{0.45\textwidth}{!}{\includegraphics*{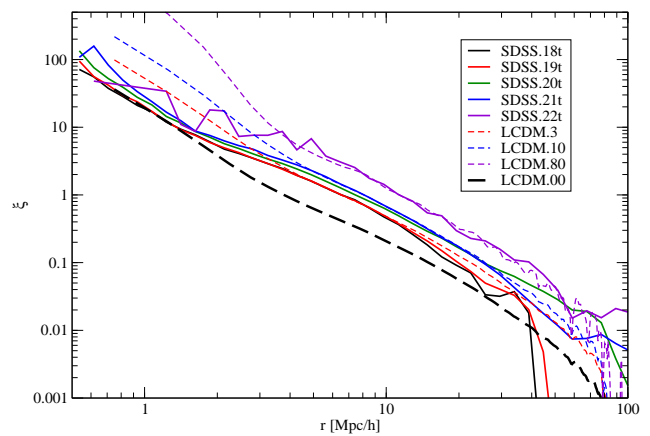}}
\caption{Correlation functions of SDSS galaxies compared with the reduced
  correlation functions of LCDM samples, using a reduction factor of 0.65.}
\label{fig:Fig10} 
\end{figure}

Fig.~\ref{fig:Fig10} shows the correlation functions of the SDSS samples
together with the reduced correlation functions of the biased model samples
LCDM.3, LCDM.10, and LCDM.80, as well as the correlation function of the whole
matter LCDM.00, shown in the figure by the black bold dashed line.  At
medium and large scales, $r \ge 3$~ \Mpc, the correlation 
function of the SDSS.22t sample is very close to the correlation
function of the biased model sample LCDM.80.  Similarly, the correlation
functions of the SDSS.20t and SDSS.21t samples are close to the correlation
function of the LCDM.10 sample, and the correlation functions of the SDSS.18t
and SDSS.19t samples are close to the correlation function of the LCDM.3
sample.  The correlation functions of last two SDSS samples bend down at
a distance $r \ge 40$~\Mpc, and deviate on larger scales from those of the LCDM
samples.  This deviation is due to smaller volumes in the SDSS.18 and
SDSS.19 samples, similarly to the behaviour of the correlation functions
of the EAGLE simulation galaxies.

On smaller scales $r \le 3$~\Mpc, the\ correlation functions of the LCDM
samples have a higher amplitude than the correlation functions of
the corresponding SDSS samples. As discussed above, in this distance range, the
correlation functions of the LCDM models reflect the internal structure of
the DM halos.  Inside the DM halos, the density of the DM particles is very high,
and all DM particles are present in our samples.  This explains the
rise of the correlation function of the LCDM samples.  The number density
of visible galaxies in our luminosity-limited galaxy samples does not
follow this tendency because most satellite galaxies and cluster
members are fainter than the luminosity limit.

\section{Fractal properties of the cosmic web}

In this section we discuss the fractal properties of the cosmic web as
measured by the correlation function and its derivatives.  The LCDM
samples differ from the SDSS, Millennium, and EAGLE samples in one
principal factor: in the LCDM samples, individual DM particles are
present, but in real and simulated galaxy samples, only galaxies brighter
than a certain limit are present.  This difference affects the
correlation functions and their derivatives, the structure and
gradient 
functions.  The largest difference is in the fractal dimension functions,
which contain information on the distribution of the DM particles
within halos, as well as on the distribution of the DM halos and galaxies
in the cosmic web.

\subsection{Fractal dimension properties of the LCDM models}

One aspect of the structure of the cosmic web is its fractal
character.  The fractal geometry of various objects in nature was
discussed by \citet{Mandelbrot:1982uq}.  The study of fractal
properties of the cosmic web has a long history.  The first essential
step for studying fractal properties of the cosmic web was made by
\citet{Soneira:1978fk}.  The authors constructed a fractal model that
allowed reproducing the angular correlation function of the Lick
catalogue of galaxies, as found by \citet{Seldner:1977aa}. Over a wide
range of mutual separations (distances), $0.01 \le r \le 10$~\Mpc, the
correlation function can be represented by a power law,
$\xi(r) = (r/r_0)^\gamma$, where $r_0 =4.5 \pm 0.5$~\Mpc\ is the
correlation length, $\gamma = -1.77$ is a characteristic power index,
and $D= 3+\gamma = 1.23 \pm 0.04$ is the fractal dimension of the
sample \citep{Peebles:1989aa, Peebles:1998aa}.

Subsequent studies have shown that the cosmic web has a
multifractal character, as found by numerous authors, starting from
\citet{Jones:1988aa}, and confirmed by \citet{Balian:1988aa},
\citet{Martinez:1990aa}, \citet{Einasto:1991oy}, \citet{Guzzo:1991aa},
\citet{Colombi:1992aa} \citet{Martinez:1993aa},
\citet{Borgani:1995aa}, \citet{Mandelbrot:1997aa},
\citet{Baryshev:1998aa}, \citet{Gaite:1999aa}, \citet{Bagla:2008aa},
\citet{Chacon-Cardona:2012aa}, and \citet{Gaite:2018aa, Gaite:2019aa}.  In
these pioneering studies, the fractal character of the web was studied
mostly locally in samples of different sizes, as discussed by
\citet{Martinez:2002fu}.

\citet{Einasto:1991ab, Einasto:1992aa} and \citet{Guzzo:1991aa} showed
that at separations $r < 3$~\Mpc,\ the correlation function of galaxies
reflects the galaxy distrubution in groups and clusters, and at
larger distances, the intracluster distribution (up to $\sim
30$~\Mpc). This was confirmed by \citet{Berlind:2002aa},
\citet{Zehavi:2004aa}, and \citet{Kravtsov:2004aa}, who noted that at
small mutual separations $r,$ the correlation function characterises
the distribution of matter within DM halos, and on larger separations,
the distribution of halos.  In the
small separation region, the spatial correlation function of the LCDM model
samples is proportional to the density of matter, and measures the
density profile inside the DM halos.  The upper left panel of
Fig.~\ref{fig:Fig3} shows that the fractal dimension function of the LCDM
samples strongly depends on the particle density limit $\rho_0$ that is used
in selecting particles for the sample.

All LCDM samples at a distance $r=0.5$~\Mpc\  have  an identical value of
the gradient function, $\gamma(0.5) = -1.5$, and
of the respective local fractal dimension, $D(0.5)=1.5$. At a distance of about
2~\Mpc, the   gradients have a minimum that depends on the particle density
limit $\rho_0$ of the samples, see Table~\ref{Tab5}.  After the minimum,
the gradient  function increases and reaches the
expected value $\gamma(100) \approx 0$ ($D(100) \approx 3.0$) for
all samples at the largest distance.

{\scriptsize 
\begin{table}[h] 
\caption{Minima  of the gradient  functions of the LCDM models.} 
\begin{tabular}{lrccr}
\hline  \hline
  Model    &$\rho_0$&$r_{\mathrm{min}}$~\Mpc&$\gamma(r_{\mathrm{min}})$ & $D(r_{\mathrm{min}})$\\
\hline  \\
LCDM.00 &0  & 1.41& $ -2.08$ & 0.92 \\ 
LCDM.10 &10 &  2.08& $-2.60$& 0.40\\
LCDM.20 &20  & 2.25& $-3.01$& $-0.01$\\
LCDM.50 &50& 2.25 & $-3.77$& $-0.77$\\
LCDM.100 &100& 2.25&$-4.54$&$-1.54$\\
\label{Tab5}                         
\end{tabular} 
\end{table} 
}

The identical values of  the gradient functions  at
$r=0.5$~\Mpc, followed by a minimum at $r \approx 2$~\Mpc, depending
on the particle density limit of the LCDM samples, can be explained by the
internal structure of the DM halos.  It is well known that DM halos have
almost identical DM density profiles, which can be described by the
NFW \citep{Navarro:1996aa} and by the \citet{Einasto:1965aa} profile.
As shown by \citet{Wang:2019aa}, the density of DM halos is better
fitted by the Einasto profile,
\begin{equation}
  \rho(r) = \rho_{-2} \exp[-2\,\alpha^{-1} ((r/r_{-2})^\alpha -1)],
\end{equation}
where $r_{-2}$ is the radius at which the logarithmic slope is $-2$,
and $\alpha$ is a shape parameter.  \citet{Wang:2019aa} showed that
density profiles of halos of very different mass are almost identical
over a very wide range of halo masses, and have the shape parameter
value $\alpha = 0.16$.  The authors calculated the logarithmic density
slope $d \log \rho/d \log r$ of the density profile over a distance
range $0.05 \le r/r_{-2} \le 30$.  Near the halo centre,
$ r/r_{-2} = 0.2$, the logarithmic slope is
$d \log \rho/d \log r = -1.5$, and it decreases to
$d \log \rho /d \log r =-3.0$ at $ r/r_{-2} = 10$ near the outer
boundary of the halo.

At small distances, the correlation functions are the mean values of the sums of
the particle mutual distances within the DM halos, averaged over the whole
volume of the samples. At the very centre, all DM halos have identical density
profiles, which explains the constant value, $\gamma(0.5) = -1.5$.  The DM
halos of different mass have different sizes, thus the outer boundaries of the halos
occur at various distances $r$, depending on the halo mass. 
In the full sample LCDM.00, small DM halos dominate,
and the minimum of the gradient  function $\gamma(r)$
has a moderate depth.  With increasing particle density limit
$\rho_0$ , low-density halos are excluded from the sample and more
massive halos dominate.  This leads to an increase in depth of
the minimum of $\gamma(r)$ and $D(r)$ functions.  More massive halos
have larger radii, thus the location of the minimum of the
$\gamma(r)$ function shifts to higher distance $r$ values.
More massive halos also have a higher density, thus the minimum of the
$\gamma(r)$ function is deeper.  The $r=r_{\mathrm{min}}$ can
be considered as the effective radius of the dominating DM halos of the
sample.

After the minimum at higher $r$ values, the distribution of DM
particles outside the halos dominates. This leads to an increase in the
$\gamma(r)$ function. In these regions, the distribution of the DM
particles of the whole cosmic web determines the correlation function
and its derivatives.  {  Fig.~\ref{fig:Fig3} and
  Fig.~2 of \citet{Zehavi:2004aa}} show that the
transition from one DM halo to the general cosmic web occurs at
$r \approx 2$~\Mpc, which agrees well with the characteristic sizes of DM
halos.
We conclude that correlation functions of our biased LCDM models
describe in a specific way the internal structure of DM halos, 
and also the fractal dimension properties of the whole cosmic web.

\subsection{Fractal dimension properties  of the galaxy samples}

The SDSS samples have at low $r$ values a large scatter of $\gamma(r)$ and
$D(r)$ functions.  The mean value of the fractal dimension function {
  at $r \approx 0.2$} is about 1.5, which is similar to the value for the LCDM
samples.  With increasing distance $r,$ the $\gamma(r)$ and $D(r)$
functions decrease and have a minimum at about
$r_{\mathrm{min}} \approx 0.8$~\Mpc. The minimum is deeper for
the brightest galaxies of sample SDSS.21. Thereafter, with increasing
$r,$ the $\gamma(r)$ and $D(r)$ functions increase rapidly up to
$D(3) \approx 2$ ($\gamma(3)=-1$), which corresponds to the fractal
dimensions of sheets.  At a still larger distance between galaxy pairs
$r,$ the $\gamma(r)$ function smoothly approaches a value
$\gamma(r) =0$ at large $r$, as expected for a random distribution of
galaxies with a fractal dimension $D =3$.

The fractal functions of the Millennium and EAGLE simulation samples start at
low $r$-values at $\gamma(r) \approx -1.5$, $D(r) \approx 1.5$, also
with a large scatter.  For
the two simulated galaxy samples, the  fractal dimension functions have a
minimum at $r_{\mathrm{min}} \approx 1.5$~\Mpc.  The minimum is deeper for brighter
galaxies.  A gradual increase in fractal dimension up
to $D=3$ at large distance follows.

The fractal dimension functions of the LCDM and galaxy samples differ in two 
important details: (i) at small distances, $r \approx 0.5$~\Mpc, the LCDM
samples have an identical  gradient, $\gamma(0.5) = -1.5$, all galaxy
samples in this distance region have large scatter of the gradient;
(ii)   the transition from { halos (clusters)} to the web 
 occurs at different scales, $r_{\mathrm{min}} \approx 2$~\Mpc\
for the LCDM models, $r_{\mathrm{min}} \approx 0.8$~\Mpc\ for real galaxies, and
$r_{\mathrm{min}} \approx 1.5$~\Mpc\ for simulated galaxies.  These small differences
show that the internal structure of the DM halos of the LCDM models differs
from the internal structure of galaxy clusters.  In our biased
LCDM model samples, all DM particles with density labels
$\rho \ge \rho_0$ are present.  Thus we see the whole density
profile of the halos up to the outer boundary of the halo.  In the real
and simulated galaxy samples, only galaxies 
brighter than the selection limit are present.  This means that in
most luminous galaxy samples, only one or a few brightest galaxies are
located within the visibility window, and the true internal structure
of clusters up to their outer boundary is invisible. This difference
is larger for real galaxies.

This result is valid not only for rich clusters of galaxies.  Ordinary
galaxies of the type of our Galaxy and M31 are surrounded by satellite
galaxies such as the Magellanic Clouds and dwarf galaxies encircling  the Galaxy,
and M32 and dwarf galaxies around M31. According to the available data, all
ordinary galaxies have such satellite systems (\citet{Einasto:1974b},
and 
  \citet{Wechsler:2018fj}). However, the luminosity of most satellites is so low
that they are invisible in galaxy surveys such as the SDSS.  For this reason,
the distribution of DM and visible matter within Galaxy-type halos is
different, as noted previously in the early study by
\citet{Einasto:1974fv}.  Groups and clusters of galaxies have more
bright member galaxies, but at large distances, fainter group members
also become invisible.  With increasing distance, the 
number of visible galaxies in clusters therefore decreases, and at a large
distance, only the brightest galaxy has a luminosity that is high enough to
fall into the visibility window of the survey, see Fig.~9 of
\citet{Tempel:2009sp}.

Cross sections of spatial 3D density fields of the SDSS and LCDM samples
are shown in Fig.~14 of \citet{Einasto:2019aa}.  Throughout the entire area
of the cross sections of the 3D fields, small isolated spots exist.
These spots correspond to primary galaxies of
groups (clusters), where satellite galaxies lie outside the visibility window, or
the number of members in groups is considerably reduced.  In the
calculation of the correlation functions, these groups enter as isolated
galaxies of the field without any nearby galaxies, or as part of groups 
with only a few member galaxies.

The difference in spatial distribution of galaxies and matter at
sub-megaparsec scales is the principal reason for the difference
between the correlation functions of the LCDM and SDSS samples at small
separations.  The visible matter in Galaxy-type halos is concentrated in
the central main galaxy. The luminosities of satellite galaxies are so
low that they contribute little to the distribution of luminous
matter. In 
contrast, DM has a smooth distribution within the halos over the whole
volume of the halo up to its outer boundary.

The correlation functions of the LCDM models with the highest particle selection
limit $\rho_0$ have a bump at $r \approx 8$~\Mpc\  that causes small
wiggles in the fractal dimension function.  Similar features are also seen
in the correlation functions of the most luminous galaxies in real and
simulated galaxy samples. Less massive { clusters (halos)} have smoother
profiles and distributions, and these features are hidden in the large
body of all galaxies.  A similar feature was detected by
\citet{Tempel:2014aa} for the correlation function of galaxies, and was
explained by a regular displacement of groups along filaments. See also
the regular location of clusters and groups in the main cluster chain
of the Perseus-Pisces supercluster \citep{Joeveer:1977py,
  Joeveer:1978b}.

\subsection{Dependence of fractal properties on the sample size}

One of the first comparisons of the spatial correlation functions of real and
model samples was made by \citet{Zeldovich:1982kl}.  Fig.~2 of this
paper compares the correlation functions of the observed sample around the
Virgo cluster with the correlation functions based on the
\citet{Soneira:1978fk} hierarchical clustering model, and with the hot dark
matter (HDM) simulation by \citet{Doroshkevich:1982fk}.  The observed
correlation function has a bend at a scale $r \approx 3$~\Mpc. On
smaller scales, the slope of the correlation function is
$\gamma(r) \approx -3$ (fractal dimension $D(r) \approx 0$), which is
also present in the HDM model.  On a larger scale, the slope is much
shallower in the observational and in the HDM model samples.  A rapid 
change in the slope of the correlation function of observational 
samples at this scale was confirmed by \citet{Einasto:1991ab,
  Einasto:1991oy}, \citet{Guzzo:1991aa}, \citet{Einasto:1992aa}, 
\citet{Einasto:1997ad}, and  \citet{Zehavi:2005aa}.  The authors
emphasised that on smaller scales, the 
slope of the structure function, $\gamma(r)$, lies between $-1.6$ to
$-2.25$, corresponding to fractal dimensions $D(r)= 1.4$ to $0.5$,
and reflecting the distribution of the galaxies in groups and clusters. At
larger distances, the slope of this function is about $-0.8$ and
reflects the inter-cluster distribution with a fractal dimension
$D \approx 2.2$.  

We here studied the spatial distribution of real and
simulated samples with sizes of 100 to 500 \Mpc.  { In
  earlier studies, samples of smaller size were available, and the
  contrast between clusters and filaments characterised the structure
  of one or a few superclusters.  In larger samples, the contrast
  between clusters and filaments is averaged over many superclusters,
  and we see the characteristic behaviour of the whole cosmic web}.

One  aspect of the fractal character is the dependence of the 
correlation length on the size of the sample.  \citet{Einasto:1986ab}
investigated the effect of voids on the distribution of galaxies
and found that the correlation length of galaxy samples increases with
the increase in size of the sample.  \citet{Pietronero:1987aa} and
\citet{Calzetti:1987aa, Calzetti:1988aa} explained that this result is
evidence for the fractal character of the galaxy 
distribution.  \citet{Martinez:2001aa} investigated the increase in
the correlation length with sample depth in more
detail. The authors found that the correlation length increases with
sample depth  up to 30~\Mpc, using a fractal that was constructed
according to the \citet{Soneira:1978fk} model.  However, in real
galaxy samples of increasing depth over 60~\Mpc,\ there is no increase in the
correlation length with depth. 

Another important aspect of the fractal character of the cosmic web is its
structure at sub-megaparsec scales, which is created by the tiny filamentary web of
the DM with sub-voids, sub-sub-voids, etc., as discussed, among others, by
\citet{Aragon-Calvo:2010ve}, \citet{Aragon-Calvo:2010wd}, and
\citet{Aragon-Calvo:2013}.  We used a spatial resolution of
the order of one megaparsec or slightly lower, therefore the fine details of the
structure of the DM filamentary web on smaller scales are invisible.

\section{Summary}

{ This paper and accompanying papers  \citep{Einasto:2019aa,
  Einasto:2020ab} address the biasing phenomenon and the multifractal
character of the cosmic web.  The biasing problem 
was recognised simultaneously with the detection of the cosmic web.
\citet{Joeveer:1977py, Joeveer:1978b} noted that ``it is very
difficult to imagine a process of galaxy and supercluster formation
which is effective enough to evacuate completely such large volumes as
cell interiors are''.  
As noted in the Introduction, there are two trends concerning the
value of the bias parameter, either $b \approx 1$, that is,, galaxies follow
matter, or $b \approx 2$ , that is, galaxies follow matter only in
over-density regions, but not in voids.  The bias value 
 $b \approx 1$ was based in most cases on measuring angular
correlation functions, $\xi(r_p,\pi)$, and integrating over the line
of sight $\pi$ to obtain the projected correlation function $w_p(r_p)$,
Eq.~(\ref{wp}). This treatment ignores the filamentary character
  of the cosmic web, however. In projection, clusters and filaments fill in
  the voids along the line of sight, which leads to the decrease in
  amplitude of the correlation functions (and power spectra), and to the
  respective bias parameters  \citep{Einasto:2020ab}.

  To estimate the large-scale bias parameter in a quantitative way, we
  used a biased $\Lambda$CDM model and calculated power spectra,
and  projected and spatial correlation functions of the biased models
  \citep{Einasto:2019aa, Einasto:2020ab}. To compare the models with
  observations, we developed the extended percolation method
  \citep{Einasto:2018aa}.  We also analysed the physical processes
  that cause the bias phenomenon \citep{Einasto:2019aa}.  The general
  results 
  of this series of studies are summarised below.

\begin{enumerate}

\item{} The combination of several physical processes (e.g.  the formation
  of halos along caustics of particle trajectories, and the phase
  synchronisation of density perturbations on various scales)
  transforms the initial random density field to the current highly
  non-random density field.

\item{} Because galaxy formation depends on the local density of
  matter,  regions exist that are devoid of galaxies. These regions fill a
  large fraction of space in the universe. This is the main reason for the rise in
  amplitude of the correlation functions and power spectra of galaxies,
  and it increases the large-scale bias parameter.

\item{} The combined evidence leads to a large-scale bias
parameter of $L_\star$ galaxies with a value $b_\star =1.85 \pm  0.15$.

\item{} We find for the correlation length of $L_\star$ galaxies the
  value of $r_0(L_\star) = 7.20 \pm 0.19$.

\end{enumerate}
}

We calculated for the biased $\Lambda$CDM models and for real and
simulated galaxy samples the spatial correlation functions of galaxies,
and their derivatives, structure functions, and fractal dimension
functions.  The correlation function and its derivatives are sensitive to
the spatial distribution of galaxies in the cosmic web as well as
to particles within DM halos. {  Our   conclusions  about the fractal
nature of the cosmic web  are listed below.}

\begin{enumerate}
  
\item{} In the biased DM models, the correlation functions at small distances (separations)
describe the distribution of particles in DM
  halos, as given by the halo density profiles.  The transition from halo
  to web properties occurs at distance $r \approx 2$~\Mpc. In 
   galaxy samples, only the brightest galaxies in clusters are
  visible, and the transition from clusters to filaments occurs at
  distance $r \approx  0.8$~\Mpc\ for real and at $r \approx  1.5$~\Mpc\ for
  simulated galaxies.  At larger separations, the correlation
  functions describe the distribution of matter and galaxies in the
  whole cosmic web.

\item{} The effective fractal dimension of the cosmic web is a
  continuous function of distances (particle and galaxy separations). At
  small separations, $ r \le 2$~\Mpc, the gradient function decreases
  from $\gamma(r) \approx -1.5$ to $\gamma(r) \approx -3$, reflecting
  the particle (galaxy)  distribution inside halos (clusters).  The minimum
  of the fractal dimension function $D(r)$ near $r \approx 2$~\Mpc\ is
  deeper for more luminous galaxies.  At medium separations,
  $2 \le r \le 10$~\Mpc, the fractal dimension grows from $\approx 0$
  to $\approx 2$ (filaments to sheets), and approaches 3 at large
  separations (random distribution).

\item{}  Real and simulated galaxies of low luminosity, $M_r \ge -19$, have
almost identical correlation lengths and amplitudes.  This is a strong
argument indicating that dwarf galaxies are satellites of brighter
galaxies, and do not form a smooth population in voids.

\end{enumerate}

\begin{acknowledgements} 

Our special thanks are to Enn Saar for providing a very efficient
code to calculate the correlation function and for many stimulating
discussions, to Ivan Suhhonenko for calculation the $\Lambda$CDM
model used in this paper, and to the anonymous referee for useful
  suggestions.  

This work was supported by institutional research fundings IUT26-2 and
IUT40-2 of the Estonian Ministry of Education and Research, and by the
Estonian Research Council grant PRG803.  We acknowledge the support by
the Centre of Excellence ``Dark side of the Universe'' (TK133)
financed by the European Union through the European Regional
Development Fund.  The study has also been supported by ICRAnet
through a professorship for Jaan Einasto.

We thank the SDSS Team for the publicly available data releases.
Funding for the SDSS and SDSS-II has been provided by the Alfred
P. Sloan Foundation, the Participating Institutions, the National
Science Foundation, the U.S. Department of Energy, the National
Aeronautics and Space Administration, the Japanese Monbukagakusho, the
Max Planck Society, and the Higher Education Funding Council for
England. The SDSS Web Site is \texttt{http://www.sdss.org/}.

The SDSS is managed by the Astrophysical Research Consortium for the 
Participating Institutions. The Participating Institutions are the 
American Museum of Natural History, Astrophysical Institute Potsdam, 
University of Basel, University of Cambridge, Case Western Reserve 
University, University of Chicago, Drexel University, Fermilab, the 
Institute for Advanced Study, the Japan Participation Group, Johns 
Hopkins University, the Joint Institute for Nuclear Astrophysics, the 
Kavli Institute for Particle Astrophysics and Cosmology, the Korean 
Scientist Group, the Chinese Academy of Sciences (LAMOST), Los Alamos 
National Laboratory, the Max-Planck-Institute for Astronomy (MPIA), 
the Max-Planck-Institute for Astrophysics (MPA), New Mexico State 
University, Ohio State University, University of Pittsburgh, 
University of Portsmouth, Princeton University, the United States 
Naval Observatory, and the University of Washington.

\end{acknowledgements}

\bibliographystyle{aa} 

\end{document}